\documentclass[11pt,a4paper]{article}

\usepackage{amsmath,amssymb,siunitx,lipsum,float,caption}
\usepackage{graphicx}
\usepackage[colorlinks=true,breaklinks=true,allcolors=blue]{hyperref}
\usepackage[numbers,sort&compress]{natbib}
\usepackage{authblk}

\graphicspath{{./imgs}}
\usepackage{enumitem}

\usepackage[margin=2.1cm]{geometry}

\DeclareMathOperator{\sign}{sign}

\begin{document}
\title{Simplified models of diffusion in radially-symmetric geometries}
\author{Luke P. Filippini, Matthew J. Simpson and Elliot J. Carr\thanks{Corresponding Author: \href{mailto:elliot.carr@qut.edu.au}{elliot.carr@qut.edu.au}}}
\affil{School of Mathematical Sciences, Queensland University of Technology, Brisbane, Australia.}

\date{}

\maketitle

\section*{Abstract} 
We consider diffusion-controlled release of particles from $d$-dimensional radially-symmetric geometries. A quantity commonly used to characterise such diffusive processes is the proportion of particles remaining within the geometry over time, denoted as $\mathcal{P}(t)$. The stochastic approach for computing $\mathcal{P}(t)$ is time-consuming and lacks analytical insight into key parameters while the continuum approach yields complicated expressions for $\mathcal{P}(t)$ that obscure the influence of key parameters and complicate the process of fitting experimental release data. In this work, to address these issues, we develop several simple surrogate models to approximate $\mathcal{P}(t)$ by matching moments with the continuum analogue of the stochastic diffusion model. Surrogate models are developed for homogeneous slab, circular, annular, spherical and spherical shell geometries with a constant particle movement probability and heterogeneous slab, circular, annular and spherical geometries, comprised of two concentric layers with different particle movement probabilities. Each model is easy to evaluate, agrees well with both stochastic and continuum calculations of $\mathcal{P}(t)$ and provides analytical insight into the key parameters of the diffusive transport system: dimension, diffusivity, geometry and boundary~conditions.

\section{Introduction} 
Mathematical modelling of diffusion-controlled transport is applied across many disciplines, including biology \cite{codling_2008, lotstedt_2015, okubo_2001}, ecology \cite{okubo_2001}, medicine \cite{arifin_2006, dash_2010, siepmann_book_2012} and physics \cite{vaccario_2015, van_kampen_2007}. Important applications include drug delivery from cylindrical \cite{jain_2022, siepmann_2012} and spherical \cite{arifin_2006, dash_2010, carr_2018, kaoui_2018, siepmann_2012} devices and the drying of fruit and vegetable products \cite{aghbashlo_2009, corzo_2008, midilli_2002, onwude_2016}. Motivated by such applications, in this paper, we explore diffusion-controlled release from $d$-dimensional radially-symmetric geometries (\hyperlink{fig1a}{Fig.~1(a)}). Here, particles diffuse within the geometry until they are absorbed at a boundary (\hyperlink{fig1b}{Fig.~1(b)}). A key quantity commonly used to characterise such diffusion processes is the proportion of particles remaining within the geometry over time, denoted as $\mathcal{P}(t)$. This quantity is equivalent to the survival probability \cite{redner_2001, simpson_2015} of an arbitrary particle and decreases over time as the number of absorbed particles increases. The shape and slope of $\mathcal{P}(t)$ (\hyperlink{fig1d}{Fig.~1(d)}) is influenced by key parameters such as the dimension, diffusivity, geometry and boundary conditions of the diffusive transport system \cite{carr_2022}.  
 
Traditionally, $\mathcal{P}(t)$ is calculated using a stochastic or continuum approach. In the stochastic approach, computing $\mathcal{P}(t)$ involves repeated simulations of a random walk model governing the motion of each individual particle. In the continuum approach, computing $\mathcal{P}(t)$ involves solving the continuum analogue of the stochastic model for the particle concentration (\hyperlink{fig1c}{Fig.~1(c)}). Both of these approaches have their drawbacks. Firstly, the stochastic approach is time-consuming and lacks analytical insight into key parameters. Secondly, the continuum approach yields complicated expressions for $\mathcal{P}(t)$ \cite{ignacio_2017, siepmann_2012} that obscure the influence of key parameters and complicate the process of fitting experimental release data~\cite{ignacio_2017, carr_2022}. To address these issues, surrogate modelling aims to develop a simplified model that accurately approximates $\mathcal{P}(t)$ and is computationally inexpensive (\hyperlink{fig1d}{Fig.~1(d)}). Previous work includes exponential, Weibull and other exponential-like models for $\mathcal{P}(t)$ and related quantities \cite{andrews_2016, simpson_2009} for slab, circular, and spherical geometries with radial symmetry \cite{arifin_2006, ignacio_2022, ignacio_2017, lotstedt_2015, onwude_2016, siepmann_2012, dash_2010}.

\begin{figure}[t!]
	\centering
	\captionsetup{justification=justified,singlelinecheck=false}
	\begin{minipage}{0.49\textwidth}\hypertarget{fig1a}{}
		\hspace*{-1.7cm}
		\centering
		\includegraphics[scale=0.7, trim = 0cm 0cm 0cm 0cm, clip]{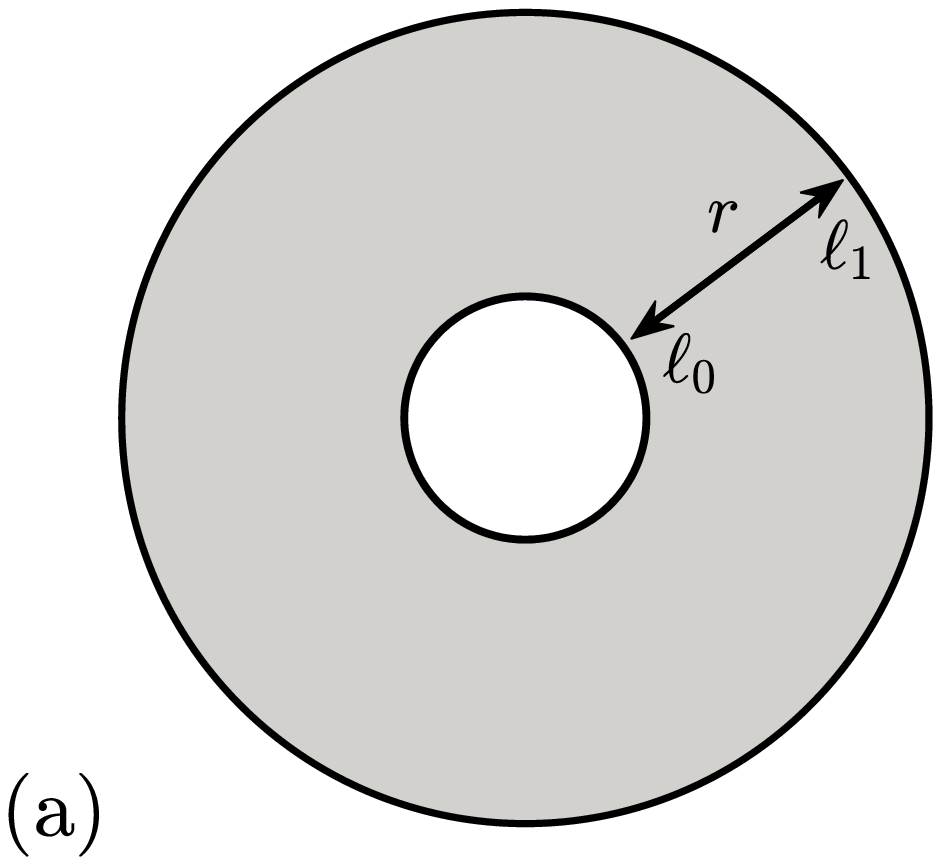}	
	\end{minipage}
	\hspace{\fill}
	\begin{minipage}{0.49\textwidth}\hypertarget{fig1b}{}
		\hspace*{-1.3cm}
		\centering
		\includegraphics[scale=0.7, trim = 0cm 0cm 0cm 0cm, clip]{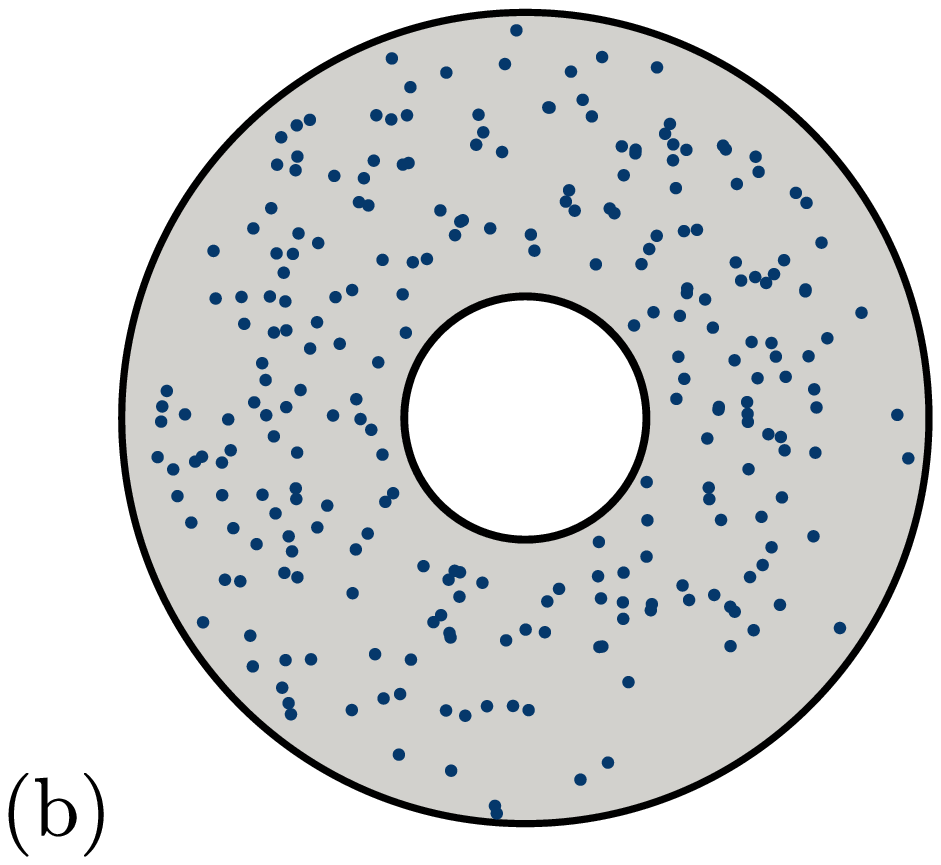}	
	\end{minipage}
	\vskip 0.2cm%
	\begin{minipage}{0.49\textwidth}\hypertarget{fib1c}{}
		\hspace*{-1.3cm}
		\centering
		\includegraphics[scale=0.23, trim = 0cm 0cm 0cm 0cm, clip]{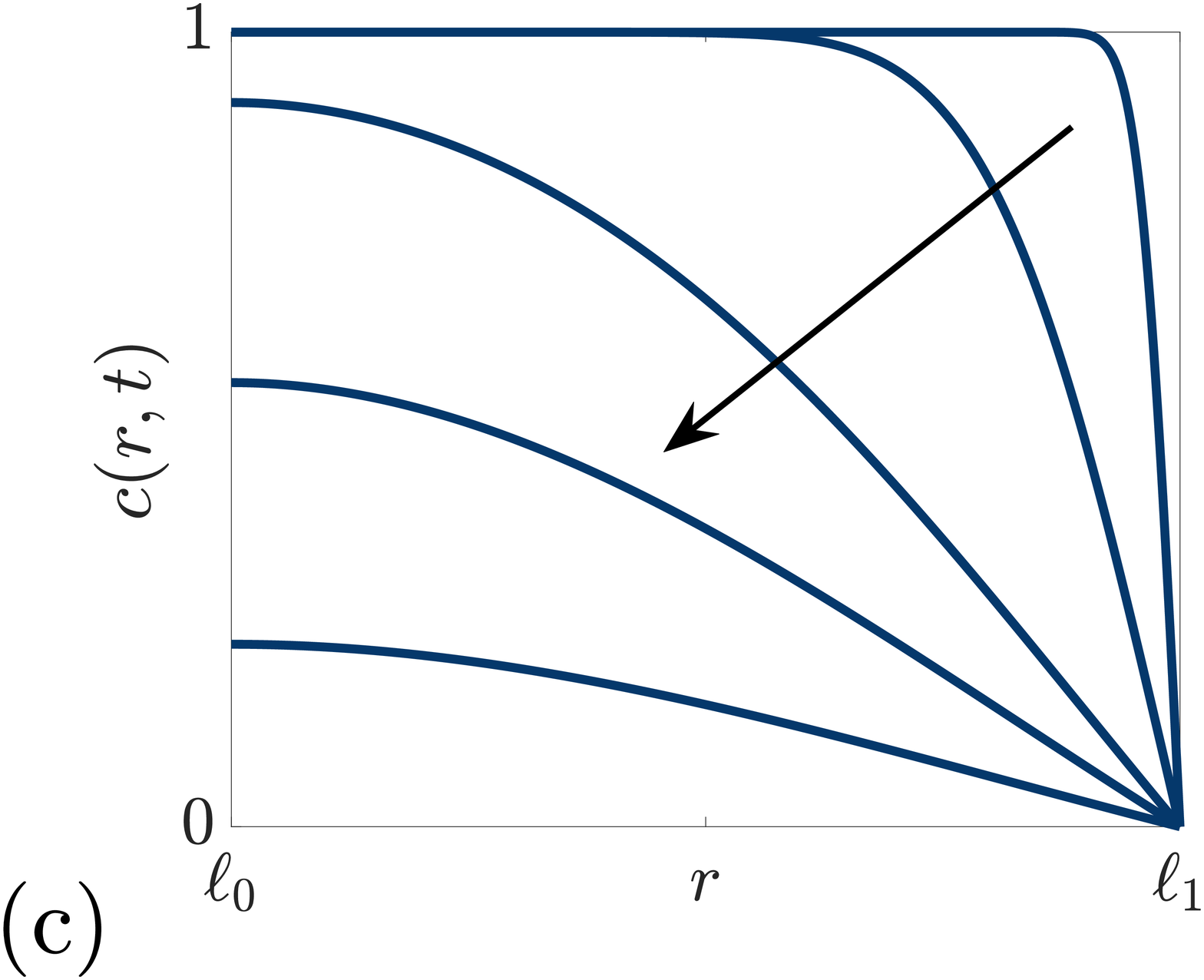}
	\end{minipage}
	\begin{minipage}{0.49\textwidth}\hypertarget{fig1d}{}
		\centering
		\includegraphics[scale=0.23, trim = 0cm 0cm 0cm 0cm, clip]{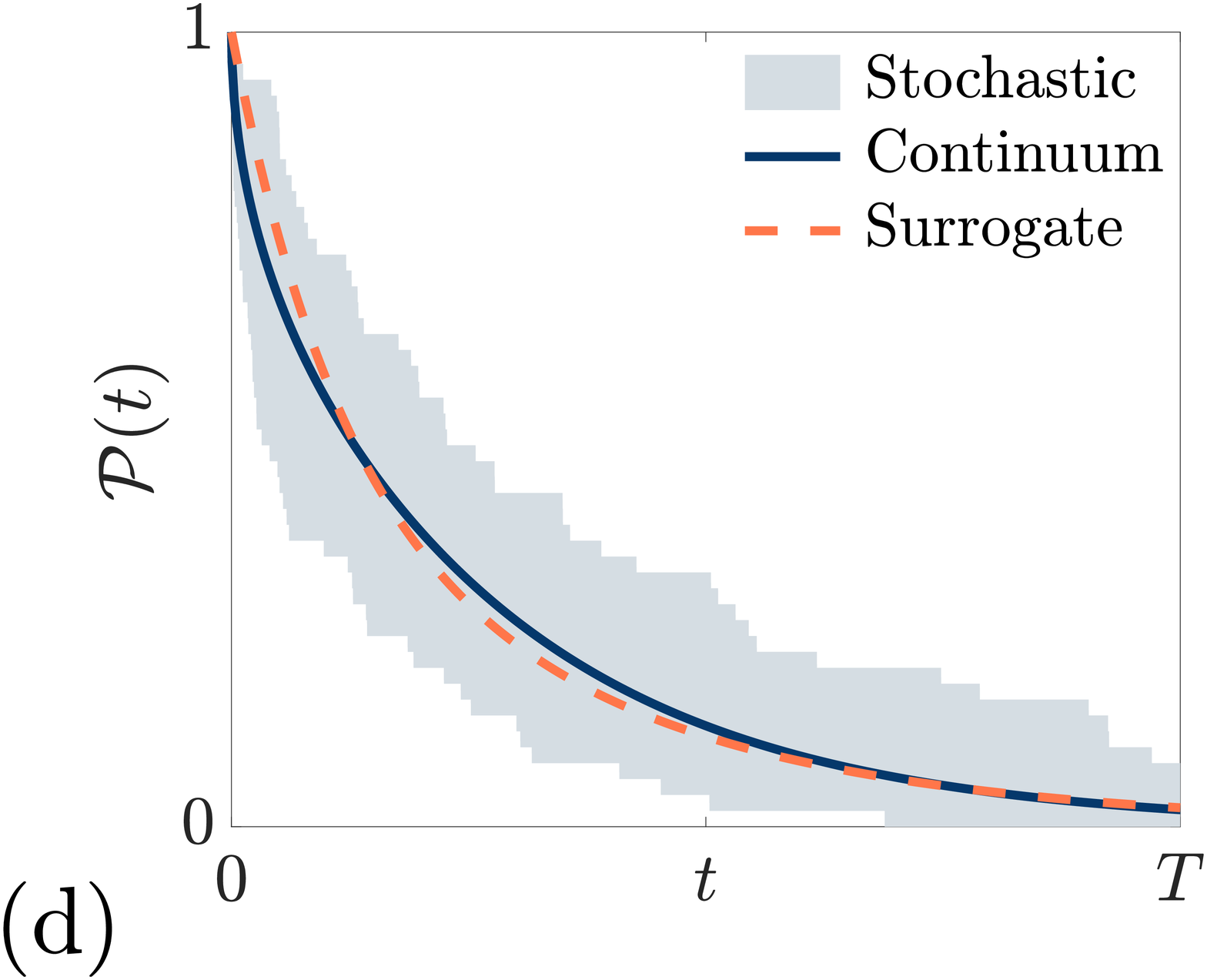}
	\end{minipage}
	\caption{(a)--(b) Diffusion-controlled release of particles from a homogeneous annular geometry with a reflecting inner boundary and absorbing outer boundary. Here, particles diffuse until they are absorbed out of the system (see section \ref{sec:stochastic}). (c) dimensionless particle concentration $c(r,t)$ obtained by solving the continuum analogue of the stochastic diffusion model (see section \ref{sec:continuum}) with the arrow indicating the direction of increasing time. (d) stochastic and continuum calculations for the proportion of particles remaining over time, $\mathcal{P}(t)$, with an example surrogate model (see sections \ref{sec:one_term}--\ref{sec:two_term_weight}) providing a simple accurate approximation to $\mathcal{P}(t)$.}
\end{figure}

In this paper, we develop several new accurate surrogate models for $\mathcal{P}(t)$ by matching moments with the continuum analogue of the stochastic diffusion model. This approach yields surrogate models that explicitly depend on, and provide analytical insight into, key parameters of the diffusive transport system. Firstly, we revisit the work of Carr \cite{carr_2022} and present one-term exponential models to approximate $\mathcal{P}(t)$, obtained by matching the zeroth moments. Secondly, we present new two-term exponential models to approximate $\mathcal{P}(t)$, obtained by matching the zeroth and first moments. Finally, we present new weighted two-term exponential models involving an arbitrary weighting of the two exponential terms, obtained by matching the zeroth, first and second moments. Our scope includes both homogeneous geometries with a constant particle movement probability and heterogeneous geometries comprised of two concentric layers with different particle movement probabilities. In addition to standard absorbing and reflecting boundary conditions, we also consider semi-absorbing boundary conditions, where particles are either absorbed or reflected with specified probabilities. Both semi-absorbing boundaries and heterogeneous geometries find application to drug delivery applications, where heterogeneous multi-layer spherical capsules encapsulated with semi-absorbing permeable outer shells are designed to better control the drug release rate \cite{carr_2018, gomes_filho_2020, kaoui_2018}. In total, we present new surrogate models for three main problems: (i) homogeneous slab, circular and spherical geometries with an absorbing or semi-absorbing boundary, (ii) homogeneous slab, annular and spherical shell geometries with absorbing, reflecting and/or semi-absorbing boundaries and (iii) heterogeneous slab, circular and spherical geometries with an absorbing or semi-absorbing boundary. Each model is easy to evaluate, agrees well with both stochastic and continuum calculations of $\mathcal{P}(t)$ and provides analytical insight into the physical parameters of the diffusive transport system: dimension, diffusivity, geometry and boundary conditions. 

The remaining sections of this work is structured as follows. Firstly, we discuss the  stochastic (section \ref{sec:stochastic}) and continuum (section \ref{sec:continuum}) models and outline how $\mathcal{P}(t)$ is calculated in each case. Secondly, we develop the new one-term (section \ref{sec:one_term}), two-term (section \ref{sec:two_term}) and weighted two-term (section \ref{sec:two_term_weight}) surrogate models for $P(t)$. Thirdly, we assess the accuracy of the surrogate models against $\mathcal{P}(t)$ obtained from the stochastic and continuum models (section \ref{sec:results}). Finally, we summarise the main elements of the work and suggest avenues for future research (section \ref{sec:conclusion}).

\section{Stochastic model}\label{sec:stochastic}
We now describe the stochastic approach for calculating $\mathcal{P}(t)$ using a random walk model for diffusive transport in $d$-dimensional radially-symmetry geometries. We consider both a homogeneous geometry ($\ell_{0}<r<\ell_{1}$) with constant particle movement probability, $P$, and a heterogeneous geometry ($\ell_{0}<r<\ell_{2}$) comprised of two concentric layers ($\ell_{0}<r<\ell_{1}$ and $\ell_{1}<r<\ell_{2}$) with different movement probabilities, $P_1$ and $P_2$. Our analysis allows for slab geometries with both an inner (left) and outer (right) boundary, circular/spherical geometries ($\ell_{0}=0$) with an outer boundary only and annular/spherical-shell geometries ($\ell_{0}>0$) with both an inner and outer boundary. 

Consider $N_{p}$ non-interacting particles and let $\mathbf{x}_j(t)$ denote the position of the $j$th particle at time $t$. Initially, the particles are uniformly distributed across the geometry:
\begin{gather}\label{eq:initDist}
	\mathbf{x}_j(0)=\begin{cases} r_j, & \text{if $d = 1$}, \\ r_j [\cos(\theta_j), \sin(\theta_j)], & \text{if $d = 2$}, \\ r_j[\cos(\theta_j)\sin(\phi_j), \sin(\theta_j)\sin(\phi_j), \cos(\phi_j)], & \text{if $d = 3$},\end{cases}
\end{gather}
where $r_j = (\ell_0^d + u_j(\ell_m^d - \ell_0^d))^{1/d}$ ($m = 1$ for homogeneous geometry and $m=2$ for heterogeneous geometry), $\phi_j = \cos^{-1}(1 - 2v_j)$, $\theta_j \sim \mathcal{U}(0,2\pi)$, $u_j \sim \mathcal{U}(0,1)$ and $v_j \sim \mathcal{U}(0,1)$ \cite{carr_2022}. Thereafter, each particle participates in a random walk with constants steps of distance $\delta > 0$ and duration $\tau>0$, where during each time step, each particle undergoes a movement or rest event with probabilities depending on the geometry under consideration.

\subsection{Homogeneous geometry}\label{sec:SRW_HOMOG}
For a homogeneous geometry, the $j$th particle moves to a new position
\begin{gather}\label{eq:nextStep}
	\mathbf{x}_j(t + \tau) = \mathbf{x}_j(t) + \begin{cases} \delta\sign(u_j - 0.5), & \text{if $d = 1$}, \\ \delta[\cos(\theta_j), \sin(\theta_j)], & \text{if $d = 2$}, \\ \delta[\cos(\theta_j)\sin(\phi_j), \sin(\theta_j)\sin(\phi_j), \cos(\phi_j)], & \text{if $d = 3$},\end{cases}
\end{gather}
with probability $P$, or remains at its current position, implying $\mathbf{x}_j(t+\tau) = \mathbf{x}_j(t)$, with probability $1 - P$. Here, $\phi_j = \cos^{-1}(1 - 2v_j)$, $\theta_j \sim \mathcal{U}(0,2\pi)$, $u_j \sim \mathcal{U}(0,1)$ and $v_j \sim \mathcal{U}(0,1)$. 

\subsection{Heterogeneous geometry}\label{sec:SRW_HETERO}
For a heterogeneous geometry, we follow \cite{carr_2020}, where the $j$th particle undergoes a movement or rest event depending on the possible new movement positions described by $\mathcal{S}_{d}(\mathbf{x}_j(t);\delta)$, the line ($d=1$), circle ($d=2$) or sphere ($d=3$) of radius $\delta$ centred on $\mathbf{x}_j(t)$. During each time step, there are three possibilities.
\begin{enumerate}[leftmargin=2em]
\item If $\mathcal{S}_{d}(\mathbf{x}_j(t);\delta)$ does not intersect the interface ($r=\ell_{1}$) and $\mathbf{x}_j(t)$ is located in the inner layer ($\ell_{0}<r<\ell_{1}$), then the $j$th particle moves to a new position (\ref{eq:nextStep}) with probability $P_1$ or remains~at its current position with probability $1-P_1$. 
\item If $\mathcal{S}_{d}(\mathbf{x}_j(t);\delta)$ does not intersect the interface ($r=\ell_{1}$) and $\mathbf{x}_j(t)$ is located in the outer layer ($\ell_{1}<r<\ell_{2}$) then the $j$th particle moves to a new position (\ref{eq:nextStep}) with probability $P_2$ or remains at its current position with probability $1-P_2$.
\item If $\mathcal{S}_{d}(\mathbf{x}_j(t);\delta)$ intersects the interface $r=\ell_1$, then the $j$th particle moves to a new position or remains at its current position with probabilities depending on the dimension $d$. For $d=1$, the $j$th particle moves to a new position
\begin{gather*}
	\mathbf{x}_j(t + \tau) = \mathbf{x}_j(t) +
	\begin{cases}
		-\delta, & \text{with probability $\mathcal{P}_1/2$}, \\
		\delta, & \text{with probability $\mathcal{P}_2/2$},
	\end{cases}
\end{gather*}
or remains at its current position with probability $1 - \mathcal{P}_1/2 - \mathcal{P}_2/2$. Here, $\mathcal{P}_k$ is the probability associated with the layer in which the position $\mathbf{x}_j(t) + \delta/2(-1)^k$ is located. For $d=2$, the $j$th particle moves to a new position
\begin{gather*}
	\mathbf{x}_j(t + \tau) = \mathbf{x}_j(t) +
	\begin{cases}
		\delta[\cos(\theta_1),\sin(\theta_1)], & \text{with probability $\mathcal{P}_1/n$}, \\
		\delta[\cos(\theta_2),\sin(\theta_2)], & \text{with probability $\mathcal{P}_2/n$}, \\
		\hspace{1.5cm} \vdots & \hspace{1.5cm} \vdots \\
		\delta[\cos(\theta_n),\sin(\theta_n)], & \text{with probability $\mathcal{P}_n/n$},
	\end{cases}
\end{gather*}
or remains at its current position with probability $1 - \sum_{k=1}^n \mathcal{P}_k/n$. Here, $n$ is a specified integer (see section \ref{sec:results}), $\theta_k = 2\pi(k-1)/n$ and $\mathcal{P}_k$ is the probability associated with the layer in which the position $\mathbf{x}_j(t) + \delta/2[\cos(\theta_k),\sin(\theta_k)]$ is located \cite{carr_2020}. For $d=3$, the $j$th particle moves to a new position
\begin{gather*}
	\mathbf{x}_j(t + \tau) = \mathbf{x}_j(t) +
	\begin{cases}
		\delta[\cos(\theta_1)\sin(\phi_1),\sin(\theta_1)\sin(\phi_1),\cos(\phi_1)], & \text{with probability $\mathcal{P}_{1,1}/n$}, \\
		\delta[\cos(\theta_1)\sin(\phi_2),\sin(\theta_1)\sin(\phi_2),\cos(\phi_2)], & \text{with probability $\mathcal{P}_{1,2}/n$}, \\
		\hspace{3cm} \vdots & \hspace{1.5cm} \vdots \\
		\delta[\cos(\theta_1)\sin(\phi_{n_2}),\sin(\theta_1)\sin(\phi_{n_2}),\cos(\phi_{n_2})], & \text{with probability $\mathcal{P}_{1,n_2}/n$}, \\
		\delta[\cos(\theta_2)\sin(\phi_1),\sin(\theta_2)\sin(\phi_1),\cos(\phi_1)], & \text{with probability $\mathcal{P}_{2,1}/n$}, \\
		\hspace{3cm} \vdots & \hspace{1.5cm} \vdots \\
		\delta[\cos(\theta_{n_1})\sin(\phi_{n_2}),\sin(\theta_{n_1})\sin(\phi_{n_2}),\cos(\phi_{n_2})], & \text{with probability $\mathcal{P}_{n_1,n_2}/n$},
	\end{cases}
\end{gather*}
or remains at its current position with probability $1 - \sum_{k=1}^{n_1}\sum_{m=1}^{n_2} \mathcal{P}_{k,m}/n$, where $n=n_1n_2$. Here, $n_{1}$ and $n_{2}$ are specified integers (see section \ref{sec:results}), $\theta_k=2\pi(k-1)/n_1$, $\phi_m = \cos^{-1}(1-2(m-1)/n_2)$ and $\mathcal{P}_{k,m}$ is the probability associated with the layer in which the position $\mathbf{x}_j(t) + \delta/2\left[\cos(\theta_k)\sin(\phi_m),\cos(\theta_k)\sin(\phi_m),\cos(\phi_m)\right]$ is located \cite{carr_2020}. 
\end{enumerate}

\subsection{Boundary conditions}
In our stochastic model, boundaries are designated as absorbing, reflecting or semi-absorbing. If a particle attempts to pass through an absorbing boundary, it is removed from the system, whereas, if it attempts to pass through a reflecting boundary, it is returned to its previous position, implying $\mathbf{x}_j(t+\tau)=\mathbf{x}_j(t)$. On the other hand, if a particle attempts to pass through a semi-absorbing inner boundary, it is absorbed with probability $P_{\rm{I}}$ and reflected with probability $1-P_{\rm{I}}$, while if a particle attempts to pass through a semi-absorbing outer boundary, it is absorbed with probability $P_{\rm{O}}$ and reflected with probability $1-P_{\rm{O}}$ \cite{erban_2007,carr_2022}. 

\subsection{Calculation of $\mathcal{P}(t)$}
For the stochastic model, $\mathcal{P}(t)$ is defined as \cite{gomes_filho_2020}
\begin{gather}\label{eq:avg_stoch}
	\mathcal{P}_s(t) = \frac{N(t)}{N_p},
\end{gather}
where $N(t)$ is the number of particles in the system at time $t$. 

\section{Continuum model}\label{sec:continuum}
We now describe the continuum approach for calculating $\mathcal{P}(t)$ using the continuum analogue of the stochastic model. For both the homogeneous and heterogeneous geometries, the continuum model takes the form of an initial-boundary value problem for the dimensionless particle concentration, $c(r,t)$, and is a valid approximation of the stochastic model in the regime of small $\delta$ and $\tau$ \cite{codling_2008, okubo_2001}. 

\subsection{Homogeneous geometry}
For the homogeneous geometry, $c(r,t)$ satisfies the $d$-dimensional radially-symmetric diffusion equation~\cite{crank_1975, ignacio_2017, okubo_2001, redner_2001, simon_2016, vaccario_2015},
\begin{gather}\label{eq:PDE_HOMOG}
	\frac{\partial c}{\partial t} = \frac{D}{r^{d-1}}\frac{\partial}{\partial r}\left(r^{d-1}\frac{\partial c}{\partial r}\right), \quad \ell_0 < r < \ell_1, \quad t > 0,
\end{gather}
subject to the initial and boundary conditions,
\begin{gather}\label{eq:IC_HOMOG}
	c(r,0) = 1, \quad \ell_0 \leq r \leq \ell_1, \\
	a_0c(\ell_0,t) - b_0\frac{\partial c}{\partial r}(\ell_0,t) = 0, \quad t > 0, \label{eq:BC1_HOMOG}\\
	a_1c(\ell_1,t) + b_1\frac{\partial c}{\partial r}(\ell_1,t) = 0, \quad t > 0 \label{eq:BC2_HOMOG},
\end{gather}
where $D = P\delta^2/(2d\tau)$ is the diffusivity. Note that $c(r,t) = \widetilde{c}(r,t)/\widetilde{c}_0$, where the particle concentration $\widetilde{c}(r,t)$ is initially uniform, $\widetilde{c}(r,0) = \widetilde{c}_0$. Here, $\widetilde{c}(r,t)$ and $\widetilde{c}_0$ are dimensional quantities that represent the number of particles and initial number of particles per unit length/area/volume \cite{carr_2022}.

\subsection{Heterogeneous geometry}
For the heterogeneous geometry, the dimensionless particle concentration is a piecewise function
\begin{gather}
c(r,t) = \begin{cases} c_{1}(r,t), & \ell_{0} \leq r \leq \ell_{1},\\
c_{2}(r,t), & \ell_{1} \leq r \leq \ell_{2},
\end{cases}
\end{gather}
where $c_{1}(r,t)$ and $c_{2}(r,t)$ satisfy the $d$-dimensional radially-symmetric diffusion equation in the inner and outer layers respectively \cite{carr_2016, carr_2018, hickson_2009, kaoui_2018}
\begin{gather}
	\frac{\partial c_1}{\partial t} = \frac{D_1}{r^{d-1}}\frac{\partial}{\partial r}\left(r^{d-1}\frac{\partial c_1}{\partial r}\right), \quad \ell_0 < r < \ell_1, \quad t > 0, \label{eq:PDE_HETERO}\\
	\frac{\partial c_2}{\partial t} = \frac{D_2}{r^{d-1}}\frac{\partial}{\partial r}\left(r^{d-1}\frac{\partial c_2}{\partial r}\right), \quad \ell_1 < r < \ell_2, \quad t > 0,
\end{gather}
subject to the initial, boundary and interface conditions,
\begin{gather}
	c_1(r,0) = 1, \quad \ell_0 \leq r \leq \ell_1, \quad
	c_2(r,0) = 1, \quad \ell_1 \leq r \leq \ell_2, \label{eq:IC_HETERO}\\
	a_0c_1(\ell_0,t) - b_0\frac{\partial c_1}{\partial r}(\ell_0,t) = 0, \quad t > 0, \label{eq:BC1_HETERO}\\
	a_1c_2(\ell_2,t) + b_1\frac{\partial c_2}{\partial r}(\ell_2,t) = 0, \quad t > 0, \label{eq:BC2_HETERO}\\
	c_1(\ell_1,t) = c_2(\ell_1,t), \quad  t > 0, \label{eq:IF1} \\
	D_1\frac{\partial c_1}{\partial r}(\ell_1,t) = D_2\frac{\partial c_2}{\partial r}(\ell_1,t), \quad t > 0. \label{eq:IF2}
\end{gather}
Here, $D_1 = P_1\delta^2/(2d\tau)$ and $D_2 = P_2\delta^2/(2d\tau)$ are the diffusivities for the inner and outer layers, respectively. The interface conditions (\ref{eq:IF1}) and (\ref{eq:IF2}) specify continuity of concentration and flux at the interface, which assumes perfect contact between the layers \cite{carr_2016, hickson_2009}. Note that $c_1(r,t) = \widetilde{c}_1(r,t)/\widetilde{c}_0$ and $c_2(r,t) = \widetilde{c}_2(r,t)/\widetilde{c}_0$, where the quantities $\widetilde{c}_1(r,t)$ and $\widetilde{c}_2(r,t)$ represent the number of particles per unit length/area/volume in the inner and outer layers, respectively.

\subsection{Boundary conditions}
The coefficients in the boundary conditions (\ref{eq:BC1_HOMOG})--(\ref{eq:BC2_HOMOG}) and (\ref{eq:BC1_HETERO})--(\ref{eq:BC2_HETERO}) depend on whether the boundaries are absorbing, reflecting or semi-absorbing:
\begin{gather}
	[a_0,b_0] = \begin{cases}[1,0], & \text{if the inner boundary is absorbing}, \\ [0,1], & \text{if the inner boundary is reflecting}, \\ [1,\beta_0], & \text{if the inner boundary is semi-absorbing},\end{cases}\\
	[a_1,b_1] = \begin{cases}[1,0], & \text{if the outer boundary is absorbing}, \\ [0,1], & \text{if the outer boundary is reflecting}, \\ [1,\beta_1], & \text{if the outer boundary is semi-absorbing}, \end{cases}
\end{gather}
with $\beta_0 = \delta/P_{\rm{I}}$ and $\beta_1 = \delta/P_{\rm{O}}$ \cite{carr_2022}. Note that for the case of the circular or spherical geometry with no inner boundary ($\ell_{0}=0$), we set $[a_0,b_0] = [0,1]$ for radial symmetry at the origin.

\subsection{Calculation of $\mathcal{P}(t)$}
For both the homogeneous continuum model (\ref{eq:PDE_HOMOG})--(\ref{eq:BC2_HOMOG}) and the heterogeneous continuum model (\ref{eq:PDE_HETERO})--(\ref{eq:IF2}), $\mathcal{P}(t)$ is defined as \cite{ignacio_2021,carr_2022}
\begin{gather*}
	\mathcal{P}_c(t) = \frac{\int_{\Omega_d} c(r,t) \, {\rm d}V}{\int_{\Omega_d} c(r,0) \, {\rm d}V},
\end{gather*}
where $\Omega_{1} = \{\mathbf{x}\in\mathbb{R}\,|\,\ell_{0}<x<\ell_{m}\}$ and $\Omega_{d} = \{\mathbf{x}\in\mathbb{R}^{d}\,|\,\ell_{0}<\|\mathbf{x}\|_{2}<\ell_{m}\}$ for $d=2,3$ ($m = 1$ for homogeneous geometry and $m=2$ for heterogeneous geometry). Using radial symmetry and the constant initial conditions (\ref{eq:IC_HOMOG}) and (\ref{eq:IC_HETERO}), $\mathcal{P}_{c}(t)$ simplifies to \cite{carr_2022,ignacio_2021}:
\begin{gather}\label{eq:avg_homog}
	\mathcal{P}_c(t) = \frac{d}{\ell_1^d - \ell_0^d}\int_{\ell_0}^{\ell_1} r^{d-1} c(r,t) \, {\rm d}r,\\\label{eq:avg_hetero}
	\mathcal{P}_c(t) = \frac{d}{\ell_2^d - \ell_0^d}\left[\int_{\ell_0}^{\ell_1} r^{d-1} c_1(r,t) \, {\rm d}r + \int_{\ell_1}^{\ell_2} r^{d-1} c_2(r,t) \, {\rm d}r\right],
\end{gather}
for the homogeneous and heterogeneous geometries, respectively. 

Clearly, calculating $\mathcal{P}_{c}(t)$ requires solving the homogeneous continuum model (\ref{eq:PDE_HOMOG})--(\ref{eq:BC2_HOMOG}) and heterogeneous continuum model (\ref{eq:PDE_HETERO})--(\ref{eq:IF2}). Alternatively, one may think of applying the averaging operators (\ref{eq:avg_homog}) or (\ref{eq:avg_hetero}) to the homogeneous or heterogeneous continuum model to derive an initial value problem for $\mathcal{P}_c(t)$. Unfortunately, this initial value problem involves $c(r,t)$ itself except for the special case of reflecting boundary conditions, where trivially, $\mathcal{P}_c(t) = 1$ for all time as no particles exit the system~\cite{carr_2022}.

\section{Surrogate models}
\subsection{Motivation}
Exact expressions for $\mathcal{P}_{c}(t)$ can be obtained by solving the homogeneous continuum model (\ref{eq:PDE_HOMOG})--(\ref{eq:BC2_HOMOG}) or heterogeneous continuum model (\ref{eq:PDE_HETERO})--(\ref{eq:IF2}) using separation of variables \cite{carr_2016,hickson_2009,kaoui_2018} and then averaging the solution by applying (\ref{eq:avg_homog}) or (\ref{eq:avg_hetero}). For example, for the case of a homogeneous disc ($d=2$) with $\ell_0=0$, $\ell_1=L$, radial symmetry at the origin ($[a_0,b_0]=[0,1]$) and a semi-absorbing boundary ($[a_1,b_1]=[1,\beta_1]$), we obtain
\begin{gather}\label{eq:exact}
	\mathcal{P}_c(t) = \frac{2}{L^2}\sum_{n=1}^{\infty} \frac{[\int_0^L rJ_0(\eta_n r)\,{\rm d}r]^2}{\int_0^L rJ_0(\eta_n r)^2\,{\rm d}r}{\rm e}^{-\eta_n^2 D t},
\end{gather}
where $\eta_n$ for $n\in\mathbb{N}^{+}$ are the positive roots of the transcendental equation
\begin{gather}\label{eq:eta_roots}
	\eta_n\frac{J_1(\eta_n L)}{J_0(\eta_n L)} = \frac{1}{\beta_1},
\end{gather}
and $J_{\nu}(\cdot)$ is the Bessel function of the first kind of order $\nu$. The problem, however, is that (\ref{eq:exact}) takes the form of an infinite series of exponential terms with complicated coefficients and the values of $\eta_n$ have to be determined numerically since closed-form expressions for the roots of (\ref{eq:eta_roots}) are not able to be determined. Moreover, to achieve sufficient accuracy for small values of time, a large number of terms need to be taken in the series (\ref{eq:exact}). All of these issues complicate both fitting experimental release data and interpreting the effect of known physical parameters, such as $L$, $D$ and $\beta_1$, on $\mathcal{P}_{c}(t)$ \cite{ignacio_2017, carr_2022}, motivating the need for surrogate modelling \cite{ignacio_2017, carr_2022}.

\subsection{Moments}\label{sec:moments}
In this work, we develop surrogate models for $\mathcal{P}(t)$ by matching moments with the continuum model. As we will see later in sections \ref{sec:one_term}--\ref{sec:two_term_weight}, this process defines surrogate models in terms of spatially-averaged moments of the continuum model. We now outline how exact expressions for these spatially-averaged moments can be calculated for both the homogeneous and heterogeneous continuum models.

\subsubsection{Homogeneous geometry}
For the homogeneous continuum model (\ref{eq:PDE_HOMOG})--(\ref{eq:BC2_HOMOG}),  the $k$th moment is defined by
\begin{gather}\label{eq:Mk}
	M_k(r) = \int_0^\infty t^k\,c(r,t) \, {\rm d}t, \quad k=0,1,2,\hdots .
\end{gather}
Closed-form solutions for $M_k(r)$ can be obtained, without prior calculation of $c(r,t)$, since $M_{k}(r)$ satisfies the differential equation \cite{carr_2019,carr_2020}: 
\begin{gather}\label{eq:Mk_ODE}
	\frac{D}{r^{d-1}}\frac{{\rm d}}{{\rm d}r}\left(r^{d-1}\frac{{\rm d}M_k}{{\rm d}r}\right) = \begin{cases} -1, & k = 0, \\ -kM_{k-1}(r), & k = 1,2,\hdots, \end{cases}
\end{gather}
subject to the boundary conditions,
\begin{gather} 
	a_0M_k(\ell_0) - b_0\frac{{\rm d}M_k}{{\rm d}r}(\ell_0) = 0, \label{eq:Mk_BC1_homog} \\
	a_1M_k(\ell_1) + b_1\frac{{\rm d}M_k}{{\rm d}r}(\ell_1) = 0. \label{eq:Mk_BC2_homog}
\end{gather}
Note that this is the same boundary value problem satisfied by the mean particle lifetime for a particle initially located at a distance $r$ from the origin \cite{redner_2001, carr_2020}. Given $M_{k}(r)$, the spatial average of the $k$th moment is then defined as
\begin{gather}\label{eq:Mk_avg_homog}
	\langle M_k(r) \rangle = \frac{d}{\ell_1^d - \ell_0^d}\int_{\ell_0}^{\ell_1} r^{d-1} M_k(r) \, {\rm d}r.
\end{gather}

\subsubsection{Heterogeneous geometry}
For the heterogeneous continuum model (\ref{eq:PDE_HETERO})--(\ref{eq:IF2}), the $k$th moment is defined by
\begin{gather}\label{eq:Mk_hetero}
	M_k(r) = 
\begin{cases}
	M_k^{(1)}(r), & \ell_0 < r < \ell_1, \\
	M_k^{(2)}(r), & \ell_1 < r < \ell_2,
\end{cases}\\
	M_k^{(1)}(r) = \int_0^\infty t^k\,c_1(r,t) \, {\rm d}t,\qquad
	M_k^{(2)}(r) = \int_0^\infty t^k\,c_2(r,t) \, {\rm d}t.
\end{gather}
Closed-form solutions for $M_k^{(1)}(r)$ and $M_k^{(2)}(r)$ can be obtained without prior calculation of $c_1(r,t)$ and $c_2(r,t)$, since $M_k^{(1)}(r)$ and $M_k^{(2)}(r)$ satisfy the differential equations \cite{carr_2019,carr_2020}:
\begin{gather}
	\frac{D_1}{r^{d-1}}\frac{{\rm d}}{{\rm d}r}\left(r^{d-1}\frac{{\rm d}M_k^{(1)}}{{\rm d}r}\right) = \begin{cases} -1, & k = 0, \\ -kM_{k-1}^{(1)}(r), & k = 1,2,\hdots, \end{cases} \label{eq:Mk1_ode} \\ \frac{D_2}{r^{d-1}}\frac{{\rm d}}{{\rm d}r}\left(r^{d-1}\frac{{\rm d}M_k^{(2)}}{{\rm d}r}\right) = \begin{cases} -1, & k = 0, \\ -kM_{k-1}^{(2)}(r), & k = 1,2,\hdots, \end{cases} \label{Mk2_ode}
\end{gather}
subject to the boundary and interface conditions
\begin{gather}
	a_0M_k^{(1)}(\ell_0) - b_0\frac{{\rm d}M_k^{(1)}}{{\rm d}r}(\ell_0) = 0, \label{eq:Mk1_BC} \\ a_1M_k^{(2)}(\ell_2) + b_1\frac{{\rm d}M_k^{(2)}}{{\rm d}r}(\ell_2) = 0, \label{eq:Mk2_BC} \\
	M_k^{(1)}(\ell_1) = M_k^{(2)}(\ell_1), \label{eq:Mk_IF1} \\ D_1\frac{{\rm d}M_k^{(1)}}{{\rm d}r}(\ell_1) = D_2\frac{{\rm d}M_k^{(2)}}{{\rm d}r}(\ell_1). \label{eq:Mk_IF2}
\end{gather}
Given $M_{k}^{(1)}(r)$ and $M_{k}^{(2)}(r)$, the spatial average of the $k$th moment is then defined as
\begin{gather}\label{eq:Mk_avg_hetero}
	\langle M_k(r) \rangle = \frac{d}{\ell_2^d - \ell_0^d}\left[\int_{\ell_0}^{\ell_1} r^{d-1}M_k^{(1)}(r) \, {\rm d}t + \int_{\ell_1}^{\ell_2} r^{d-1} M_k^{(1)}(r) \, {\rm d}t\right].
\end{gather}

\subsection{Surrogate model 1: One-term exponential model}\label{sec:one_term}
We now consider a surrogate model for $\mathcal{P}_c(t)$ consisting of a single exponential term \cite{carr_2022},
\begin{gather}\label{eq:c1}
	S_1(t) = {\rm e}^{-\lambda t},
\end{gather}
where $\lambda > 0$ is a constant which depends on the dimension, diffusivity, geometry and boundary conditions. Note that (\ref{eq:c1}) is a sensible candidate model since it agrees with $\mathcal{P}_c(t)$ at initial time ($t=0$) and has the correct limiting behaviour at large times ($t\rightarrow\infty$) (see, e.g., $\mathcal{P}_c(t)$ in equation (\ref{eq:exact})). To determine $\lambda$, we follow \cite{carr_2022} and match the zeroth moments of $S_{1}(t)$ (\ref{eq:c1}) and $\mathcal{P}_c(t)$ (\ref{eq:avg_homog}),
\begin{gather}\label{eq:zeroth}
	\int_0^\infty S_1(t) \, {\rm d}t = \int_0^\infty \mathcal{P}_c(t) \, {\rm d}t.
\end{gather}
Substituting $S_1(t)$ (\ref{eq:c1}) and $\mathcal{P}_c(t)$ ((\ref{eq:avg_homog}) or (\ref{eq:avg_hetero})) into equation (\ref{eq:zeroth}), integrating exactly on the left hand side, reversing the order of integration on the right hand side and rearranging yields
\begin{gather}\label{eq:lambdaGen}
	\lambda = \frac{1}{\langle M_0(r) \rangle},
\end{gather}
where $\langle M_0(r)\rangle$ is defined in section \ref{sec:moments}.

We now present several one-term exponential models for $\mathcal{P}(t)$. The models are developed for the seven distinct cases outlined in \hyperlink{tab1}{Table 1} involving both homogeneous (Cases A--E) and heterogeneous (Cases F--G) geometries and various combinations of boundary conditions. Each model is presented by providing a closed-form expression for $\lambda$ appearing in the one-term exponential model (\ref{eq:c1}). For the homogeneous geometries, $\lambda$ is calculated by solving the boundary value problem (\ref{eq:Mk_ODE})--(\ref{eq:Mk_BC2_homog}) for $M_0(r)$, calculating $\langle M_0(r)\rangle$ (\ref{eq:Mk_avg_homog}) and then computing $\lambda$ (\ref{eq:lambdaGen}). For the heterogeneous geometries, $\lambda$ is calculated by solving the boundary value problem (\ref{eq:Mk1_ode})--(\ref{eq:Mk_IF2}) for $M_0^{(1)}(r)$ and $M_0^{(2)}(r)$, calculating $\langle M_0(r)\rangle$ (\ref{eq:Mk_avg_hetero}) and then computing $\lambda$ (\ref{eq:lambdaGen}). For Cases C--E, we note that $\lambda$ is expressed generally for any dimension $d$ using the definite integral $\int_{\ell_{0}}^{\ell_{1}} r^{1-d}\,{\rm d}r$, which is equal to $\ell_{1}-\ell_{0}, \ln(\ell_{1}/\ell_{0}),1/\ell_{0}-1/\ell_{1}$ for $d=1,2,3$, respectively.

\begin{table}[t!]\hypertarget{tab1}{}
	\centering
	\begin{tabular}{c p{3cm} p{3cm} p{3.2cm} c c c c}
		\hline
		Case & Geometry & Inner Boundary & Outer Boundary & $a_0$ & $b_0$ & $a_1$ & $b_1$ \\
		\hline
		A & homogeneous & -- & absorbing & 0 & 1 & 1 & 0 \\
		B & homogeneous & -- & semi-absorbing & 0 & 1 & 1 & $\beta_1$ \\
		C & homogeneous & reflecting & absorbing & 0 & 1 & 1 & 0 \\
		D & homogeneous & reflecting & semi-absorbing & 0 & 1 & 1 & $\beta_1$ \\
		E & homogeneous & absorbing & absorbing & 1 & 0 & 1 & 0 \\
		F & heterogeneous & -- & absorbing & 0 & 1 & 1 & 0 \\
		G & heterogeneous & -- & semi-absorbing & 0 & 1 & 1 & $\beta_1$ \\
		\hline
	\end{tabular}
	\caption{Geometry and boundary parameters for Cases A--G. Note that for Cases A, B, F and G there is no inner boundary ($\ell_{0}=0$), so we set $[a_0,b_0] = [0,1]$ for radial symmetry at the origin.
}
\end{table}

\bigskip
\noindent\textbf{Case A}: homogeneous slab, circular or spherical geometry ($\ell_0 = 0$ and $\ell_1 = L$) with radial symmetry at the origin ($[a_0,b_0] = [0,1]$) and an absorbing outer boundary ($[a_1,b_1]=[1,0]$) 
\begin{gather}\label{eq:model1_case1}
	\lambda = \frac{d(d+2)D}{L^2}.
\end{gather}

\medskip
\noindent\textbf{Case B}: homogeneous slab, circular or spherical geometry ($\ell_0 = 0$ and $\ell_1 = L$) with radial symmetry at the origin ($[a_0,b_0] = [0,1]$) and a semi-absorbing outer boundary ($[a_1,b_1]=[1,\beta_1]$) 
\begin{gather}\label{eq:model1_case2}
	\lambda = \frac{d(d+2)D}{L^2 + \beta_1 L(d+2)}.
\end{gather}

\medskip
\noindent\textbf{Case C}: homogeneous slab, annular or spherical shell geometry ($\ell_0 > 0$) with a reflecting inner boundary ($[a_0,b_0] = [0,1]$) and an absorbing outer boundary ($[a_1,b_1] = [1,0]$)
\begin{gather}\label{eq:model1_case3}
	\lambda = \frac{d(d+2)(\ell_1^d - \ell_0^d)D}{\ell_1^{d+2} + (d+2)[\ell_0^{2d}\int_{\ell_0}^{\ell_1}r^{1-d}\,{\rm d}r-\ell_0^d(\ell_1^2-\ell_0^2)] - \ell_0^{d+2}}.
\end{gather}

\medskip
\noindent\textbf{Case D}: homogeneous slab, annular or spherical shell geometry ($\ell_0 > 0$) with a reflecting inner boundary ($[a_0,b_0] = [0,1]$) and a semi-absorbing outer boundary ($[a_1,b_1] = [1,\beta_{1}]$)
\begin{gather}\label{eq:model1_case4}
	\lambda = \frac{d(d+2)(\ell_1^d - \ell_0^d)D}{\ell_1^{d+2} + (d+2)[\ell_0^{2d}\int_{\ell_0}^{\ell_1}r^{1-d}\,{\rm d}r-\ell_0^d(\ell_1^2-\ell_0^2)+\beta_1\ell_1^{1-d}(\ell_1^d-\ell_0^d)^2] - \ell_0^{d+2}}.
\end{gather}

\medskip
\noindent\textbf{Case E}: homogeneous slab, annular or spherical shell geometry with absorbing inner ($[a_0,b_0] = [1,0]$) and outer ($[a_1,b_1] = [1,0]$) boundaries
\begin{gather}\label{eq:model1_case5}
	\lambda = \frac{4d(d+2)(\ell_1^d - \ell_0^d)D}{4(\ell_1^{d+2}-\ell_0^{d+2}) - (d+2)(\ell_1^2-\ell_0^2)^2[\int_{\ell_0}^{\ell_1}r^{1-d}\,{\rm d}r]^{-1}}.
\end{gather}

\medskip
\noindent\textbf{Case F}: heterogeneous slab, circular or spherical geometry ($\ell_0 = 0$, $\ell_2 = L$) with radial symmetry at the origin ($[a_0,b_0] = [0,1]$) and an absorbing boundary ($[a_1,b_1] = [1,0]$)
\begin{gather}\label{eq:model1_case6}
	\lambda = \frac{d(d+2) D_1 D_2}{L^2 D_1 +\ell_1^{d+2}(D_2 - D_1)/L^d }.
\end{gather}

\medskip
\noindent\textbf{Case G}: heterogeneous slab, circular or spherical geometry ($\ell_0 = 0$, $\ell_2 = L$) with radial symmetry at the origin ($[a_0,b_0] = [0,1]$) and a semi-absorbing boundary ($[a_1,b_1] = [1,\beta_{1}]$)
\begin{gather}\label{eq:model1_case7}
	\lambda = \frac{d(d+2) D_1 D_2}{(L^2 +\beta_1 L(d+2))D_1 + \ell_1^{d+2}(D_2 - D_1)/L^d }.
\end{gather}

\bigskip
\noindent The above results yield easy-to-evaluate surrogate models that provide analytical insight into the role of dimension, diffusivity, geometry and boundary conditions on the proportion of particles remaining over time, $\mathcal{P}(t)$. For Case A (\ref{eq:model1_case1}), we observe that increasing the dimension $d$, increasing the diffusivity $D$ or decreasing the radius $L$ increases the decay rate $\lambda$. For Case B (\ref{eq:model1_case2}), decreasing $\beta_{1}$ (i.e.~increasing the absorption probability $P_{\rm{O}}$) also increases the decay rate $\lambda$. All these observations make physical sense when considering the homogeneous stochastic model (section \ref{sec:stochastic}) as particles are more likely to move outward than inward when the number of dimensions $d$ is increased, particles jump more frequently or jump further when $D$ is increased, particles have have less distance to reach the absorbing boundary when $L$ is decreased and particles are more likely to be absorbed when reaching the outer boundary when $\beta_{1}$ is decreased. For Case F (\ref{eq:model1_case6}) and Case G (\ref{eq:model1_case7}), moving the interface ($r=\ell_{1}$) closer to the outer boundary ($r=L$) increases $\lambda$ if $D_{2} < D_{1}$ while moving the interface ($r=\ell_{1}$) closer to the origin ($r=0$) increases $\lambda$ if $D_{2} > D_{1}$. Both observations are consistent with the heterogeneous stochastic model (section \ref{sec:stochastic}). 

\subsection{Surrogate model 2: Two-term exponential model}\label{sec:two_term}
The one-term exponential model (\ref{eq:c1}) fails to accurately capture the fast early decay and slow late decay of $\mathcal{P}(t)$ \cite{carr_2022}. To address this, we explore a surrogate model for $\mathcal{P}_{c}(t)$ consisting of two exponential terms,
\begin{gather}\label{eq:model2}
S_2(t) = \frac{1}{2}[{\rm e}^{-\lambda_{1}t}+{\rm e}^{-\lambda_{2}t}],
\end{gather}
where $\lambda_1 > 0$ and $\lambda_2 > 0$ are constants that depend on the dimension, diffusivity, geometry and boundary conditions. The two-term exponential model (\ref{eq:model2}) represents the simplest possible extension to two exponential terms with the factor of $1/2$ ensuring that $S_2(0) = 1$.  The inclusion of a second exponential term in (\ref{eq:model2}) yields a time-dependent decay rate $\smash{\widetilde{\lambda}_{2}(t) = -S_2'(t)/S_2(t)}$, which decreases monotonically from $(\lambda_{1}+\lambda_{2})/2$ at $t = 0$ to $\min(\lambda_{1},\lambda_{2})$ as $t\rightarrow\infty$. The two-term exponential model therefore accommodates faster early decay and slower late decay that cannot be captured by the constant decay rate of the one-term exponential model (\ref{eq:c1}).

To obtain $\lambda_1$ and $\lambda_2$, we match the zeroth and first moments of $S_2(t)$ and $\mathcal{P}_c(t)$,
\begin{gather}
	\int_0^\infty S_2(t) \, {\rm d}t = \int_0^\infty \mathcal{P}_c(t) \, {\rm d}t, \label{eq:model2_zeroth} \\
	\int_0^\infty t\,S_2(t) \, {\rm d}t = \int_0^\infty t\,\mathcal{P}_c(t) \, {\rm d}t. \label{eq:model2_first}
\end{gather}
Substituting $S_2(t)$ (\ref{eq:model2}) and $\mathcal{P}_c(t)$ ((\ref{eq:avg_homog}) or (\ref{eq:avg_hetero})) into equations (\ref{eq:model2_zeroth}) and (\ref{eq:model2_first}), integrating exactly on the left hand side and reversing the order of integration on the right hand side yields
\begin{gather}
	\frac{1}{2}\left[\frac{1}{\lambda_1} + \frac{1}{\lambda_2}\right] = \langle M_0(r) \rangle, \label{eq:model2_eq1} \\
	\frac{1}{2}\left[\frac{1}{\lambda_1^2} + \frac{1}{\lambda_2^2}\right] = \langle M_1(r) \rangle, \label{eq:model2_eq2}
\end{gather}
where $\langle M_0(r)\rangle$ and $\langle M_1(r)\rangle$ are defined in section \ref{sec:moments}. The exact solution of equations (\ref{eq:model2_eq1}) and (\ref{eq:model2_eq2}) is given by
\begin{gather}\label{eq:lambda_model2}
	\lambda_{1,2} = \frac{1}{\langle M_0(r) \rangle \pm \sqrt{\langle M_1(r) \rangle - \langle M_0 (r) \rangle^2}},
\end{gather}
which is easily verified by substitution.

We now present several two-term exponential models for $\mathcal{P}(t)$. The models are again developed for the seven cases outlined in \hyperlink{tab1}{Table 1}, with each model presented by providing closed-form expressions for $\lambda_{1}$ and $\lambda_{2}$ appearing in the two-term exponential model (\ref{eq:model2}). For the homogeneous geometries (Cases A--E), $\lambda_{1}$ and $\lambda_{2}$ are calculated by solving the boundary value problem (\ref{eq:Mk_ODE})--(\ref{eq:Mk_BC2_homog}) for $k=0,1$, calculating $\langle M_k(r)\rangle$ (\ref{eq:Mk_avg_homog}) for $k=0,1$ and then computing $\lambda_{1}$ and $\lambda_{2}$ (\ref{eq:lambda_model2}). For the heterogeneous geometries (Cases F--G), $\lambda_{1}$ and $\lambda_{2}$ are calculated by solving the boundary value problem (\ref{eq:Mk1_ode})--(\ref{eq:Mk_IF2}) for $k=0,1$, calculating $\langle M_k(r)\rangle$ (\ref{eq:Mk_avg_hetero}) for $k=0,1$ and then computing $\lambda_{1}$ and $\lambda_{2}$ (\ref{eq:lambda_model2}). Results for Cases C/D and Cases F/G are combined for succinctness. In these cases, the formulas are given for Case D and G only, with the formulas for Case C and F obtained by setting $\beta_{1} = 0$. 

\bigskip
\noindent\textbf{Case A:} homogeneous slab, circular or spherical geometry ($\ell_0 = 0$, $\ell_1 = L$) with radial symmetry at the origin ($[a_0,b_0]=[0,1]$) and an absorbing boundary ($[a_1,b_1]=[1,0]$)
\begin{gather}\label{eq:model2_case1}
	\lambda_{1,2} = \frac{d(d+2)D}{L^2(1 \pm \sqrt{d/(d+4)})}.
\end{gather}

\medskip
\noindent\textbf{Case B:} homogeneous slab, circular or spherical geometry ($\ell_0 = 0$, $\ell_1 = L$) with radial symmetry at the origin ($[a_0,b_0]=[0,1]$) and a semi-absorbing boundary ($[a_1,b_1]=[1,\beta_1]$)
\begin{gather}\label{eq:model2_case2}
	\lambda_{1,2} = \frac{d(d+2)D}{L^2(1 \pm \sqrt{d/(d+4)}) + \beta_1 L(d+2)}.
\end{gather}

\medskip
\noindent\textbf{Case C/D:} homogeneous slab, annular or spherical shell geometry ($\ell_0 > 0$) with a reflecting inner boundary ($[a_0,b_0]=[0,1]$) and a semi-absorbing outer boundary ($[a_1,b_1]=[1,\beta_1]$)

\medskip
\noindent\textit{Slab $(d=1)$}
\begin{gather*}
	\lambda_{1,2}=\frac{3D}{(\ell_{1}-\ell_{0})^{2}(1\pm 1/\sqrt{5}) + 3\beta_{1}(\ell_{1}-\ell_{0})}.
\end{gather*}

\medskip
\noindent\textit{Annular $(d=2)$}
\begin{gather*}
	\lambda_{1,2}=\frac{8D(\ell_{1}^{2}-\ell_{0}^{2})}{(\ell_{1}^{2}-\ell_{0}^{2})(\ell_{1}^{2}-3\ell_{0}^{2})+4\ell_{0}^{4}\log(\ell_1/\ell_0)+4\beta_{1}(\ell_{1}^{2}-\ell_{0}^{2})^{2}/\ell_{1}\pm\sqrt{\kappa_2/3}}, \\
	\kappa_2 = (\ell_{1}^{2}-\ell_{0}^{2})^{3}(\ell_{1}^{2}-7\ell_{0}^{2})-24\ell_{0}^{4}\ell_{1}^{2}\log(\ell_1/\ell_0)(2\ell_{0}^{2}\log(\ell_1/\ell_0)+\ell_{0}^{2}-\ell_{1}^{2}).
\end{gather*}

\medskip
\noindent\textit{Spherical shell $(d=3)$}
\begin{gather*}
	\lambda_{1,2}=\frac{15D(\ell_{1}^{3}-\ell_{0}^{3})}{(\ell_{1}-\ell_{0})^{3}(\ell_{1}^{2}+3\ell_{0}\ell_{1}+6\ell_{0}^{2}+5\ell_{0}^{3}/\ell_{1}\pm\sqrt{3\kappa_3/7})+5\beta_1(\ell_{1}^{3}-\ell_{0}^{3})^{2}/\ell_{1}^{2}}, \\
	\kappa_3 = \ell_{1}^{4}+6\ell_{0}\ell_{1}^{3}+21\ell_{0}^{2}\ell_{1}^{2}+41\ell_{0}^{3}\ell_{1}+36\ell_{0}^{4}.
\end{gather*}

\medskip
\noindent\textbf{Case E:} homogeneous slab, annular or spherical shell geometry ($\ell_0 > 0$) with absorbing inner ($[a_0,b_0]=[1,0]$) and outer ($[a_1,b_1]=[1,0]$) boundaries

\medskip
\noindent\textit{Slab $(d=1)$}
\begin{gather*}
	\lambda_{1,2}=\frac{12D}{(\ell_{1}-\ell_{0})^{2}(1\pm1/\sqrt{5})}.
\end{gather*}

\medskip
\noindent\textit{Annular $(d=2)$}
\begin{gather*}
	\lambda_{1,2}=\frac{8D\log(\ell_1/\ell_0)}{(\ell_{0}^{2}+\ell_{1}^{2})\log(\ell_1/\ell_0)-(\ell_{1}^{2}-\ell_{0}^{2})\pm\sqrt{\xi_{2,1}/3}}, \\
	\xi_{2,1} = 3(\ell_{1}^{2}-\ell_{0}^{2})^{2}-3(\ell_{1}^{4}-\ell_{0}^{4})\log(\ell_1/\ell_0)+(\ell_{1}^{2}-\ell_{0}^{2})^{2}\log^{2}(\ell_1/\ell_0).
\end{gather*}
\noindent\textit{Spherical shell $(d=3)$}
\begin{gather*}
	\lambda_{1,2}=\frac{60D(\ell_{1}^{3}-\ell_{0}^{3})}{(\ell_{1}-\ell_{0})^{3}(4\ell_{0}^{2}+7\ell_{0}\ell_{1}+4\ell_{1}^{2}\pm\sqrt{3\xi_{3,1}/7})}, \\
	\xi_{3,1} = 16\ell_{1}^{4}+26\ell_{0}\ell_{1}^{3}+21\ell_{0}^{2}\ell_{1}^{2}+26\ell_{0}^{3}\ell_{1}+16\ell_{0}^{4}.
\end{gather*}

\medskip
\noindent\textbf{Case F/G:} heterogeneous slab, circular or spherical geometry ($\ell_0 = 0$, $\ell_2 = L$), radial symmetry at the origin ($[a_0,b_0]=[0,1]$) and a semi-absorbing boundary ($[a_1,b_1]=[1,\beta_1]$)
\begin{gather*}
	\lambda_{1,2}=\frac{d(d+2)D_{1}D_{2}}{[L^{2}+\beta_{1} L(d+2)]D_{1}+[\ell_{1}^{d+2}(D_{2}-D_{1})\pm\sqrt{\sigma_d/(d+4)}]/L^d}, \\
	\begin{split}
		\sigma_d &= d(d+4)(D_{2}-D_{1})D_{1}L^{d+2}\ell_{1}^{d+2}-(d+4)(D_{1}-D_{2})^{2}\ell_{1}^{2d+4}+\\&(d+2)((d+2)D_{1}^{2}-(d+4)D_{1}D_{2}+2D_{2}^{2})L^{d}\ell_{1}^{d+4}+dD_{1}^{2}L^{2d+4}.
	\end{split}
\end{gather*}

\bigskip
\noindent The above results yield easy-to-evaluate surrogate models that provide analytical insight into the role of dimension, diffusivity, geometry and boundary conditions on the proportion of particles remaining over time, $\mathcal{P}(t)$. As mentioned earlier, the two-term exponential model (\ref{eq:model2}) accommodates faster early decay and slower late decay that cannot be captured by the constant decay rate of the one-term exponential model (\ref{eq:c1}). This behaviour is clearly evident for Case A, where the expressions for $\lambda_{1}$ and $\lambda_{2}$ in the two-term exponential model (\ref{eq:model2_case1}) take a similar form to the expression for $\lambda$ in the one-term exponential model (\ref{eq:model1_case1}), with the exception of correction terms in the denominator depending on the dimension $d$. Using these expressions for $\lambda_{1}$ and $\lambda_{2}$, we see that the two-term exponential model exhibits an initial decay rate of $(\lambda_{1}+\lambda_{2})/2 = d(d+2)(d+4)D/(4L^2)$, which exceeds its late decay rate of $\lambda_{1} = d(d+2)D/[L^{2}(1+\sqrt{d/(d+4)})]$ for all $d=1,2,3$. Comparing these decay rates to the constant decay rate of $\lambda = d(d+2)D/L^2$ for the one-term exponential model (\ref{eq:model1_case1}), it is clear that the two-term exponential model exhibits a larger initial decay rate and a smaller late decay rate. For Case A, we also observe that the early decay rate for the two-term exponential model is fastest for $d=3$ and slowest for $d=1$, and the later decay rate is slowest for $d=3$ and fastest for $d=1$, both of which are consistent with the behaviour of $\mathcal{P}(t)$ \cite{carr_2022}. 

\subsection{Surrogate model 3: Weighted two-term exponential model}\label{sec:two_term_weight}
Finally, we consider a surrogate model for $\mathcal{P}(t)$ which generalizes the two-term model (\ref{eq:model2}) to an arbitrary weighting of the two exponential terms:
\begin{gather}\label{eq:model3}
S_3(t)=\theta {\rm e}^{-\lambda_{1}t}+(1-\theta){\rm e}^{-\lambda_{2}t},
\end{gather}
where $\lambda_1 > 0$, $\lambda_2 > 0$ and $\theta\in (0,1)$ are constants that depend on the dimension, diffusivity, geometry and boundary conditions. In a similar manner to the two-term exponential model (\ref{eq:model2}), the weighted two-term exponential model (\ref{eq:model3}) exhibits a time-dependent decay rate  $\smash{\widetilde{\lambda}_{3}(t)=-S_3'(t)/S_3(t)}$, however, the decay rate now decreases monotonically from $\theta\lambda_{1}+(1-\theta)\lambda_{2}$ at $t = 0$ to $\min(\lambda_{1},\lambda_{2})$ as~$t\rightarrow\infty$.

To obtain $\lambda_1$, $\lambda_2$ and $\theta$, we match the zeroth, first and second moments of $S_3(t)$ and $\mathcal{P}_c(t)$,
\begin{gather}
	\int_0^\infty S_3(t) \, {\rm d}t = \int_0^\infty \mathcal{P}_c(t) \, {\rm d}t, \label{eq:model3_zeroth} \\ 
	\int_0^\infty t \, S_3(t) \, {\rm d}t = \int_0^\infty t \, \mathcal{P}_c(t) \, {\rm d}t, \label{eq:model3_first} \\ 
	\int_0^\infty t^2 \, S_3(t) \, {\rm d}t = \int_0^\infty t^2 \, \mathcal{P}_c(t) \, {\rm d}t. \label{eq:model3_second}
\end{gather}
Substituting $S_3(t)$ (\ref{eq:model3}) and $\mathcal{P}_c(t)$ ((\ref{eq:avg_homog}) or (\ref{eq:avg_hetero})) into equations (\ref{eq:model3_zeroth})--(\ref{eq:model3_second}), integrating exactly on the left hand side and reversing the order of integration on the right hand side yields
\begin{gather}
	\frac{\theta}{\lambda_1} + \frac{1-\theta}{\lambda_2} = \langle M_0(r) \rangle, \label{eq:model3_eq1}\\
	\frac{\theta}{\lambda_1^2} + \frac{1-\theta}{\lambda_2^2} = \langle M_1(r) \rangle, \label{eq:model3_eq2}\\
	2\left[\frac{\theta}{\lambda_1^3} + \frac{1-\theta}{\lambda_2^3}\right] = \langle M_2(r) \rangle, \label{eq:model3_eq3}
\end{gather}
where $\langle M_0(r)\rangle$, $\langle M_1(r)\rangle$ and $\langle M_2(r)\rangle$ are defined in section \ref{sec:moments}. The appropriate exact solution of equations (\ref{eq:model3_eq1})--(\ref{eq:model3_eq3}) is given by
\begin{gather}
	\lambda_1 = \frac{1}{\langle M_0(r) \rangle + \sqrt{(1-\theta)[\langle M_1(r) \rangle - \langle M_0(r) \rangle^2]/\theta}}, \label{eq:lambda1_weight}\\ 
	\lambda_2 =  \frac{1}{\langle M_0(r) \rangle - \sqrt{\theta[\langle M_1(r) \rangle - \langle M_0(r) \rangle^2]/(1-\theta)}}, \label{eq:lambda2_weight} \\
	\theta = \frac{1}{2} + \frac{1}{2}\sqrt{\frac{\omega}{\omega+4}}, \label{eq:theta} \\ 
	\omega = \left[\frac{6\langle M_0(r)\rangle(\langle M_1(r)\rangle - \langle M_0(r)\rangle^2) + 2\langle M_0(r)\rangle^3 - \langle M_2(r)\rangle}{2(\langle M_1(r)\rangle - \langle M_0(r)\rangle^2)^{3/2}}\right]^2. \label{eq:omega}
\end{gather}
Note that the expressions for $\lambda_{1}$ and $\lambda_{2}$ here are different from those given for the two-term exponential model (\ref{eq:lambda_model2}) expect for the special case when $\theta=1/2$ ($\omega=0$).

We now present weighted two-term exponential models of $\mathcal{P}(t)$ for Cases A--E outlined in \hyperlink{tab1}{Table 1}. Each model is presented by providing closed-form expressions for $\lambda_{1}$, $\lambda_{2}$ and $\omega$, which when combined with the expression for $\theta$ (\ref{eq:theta}) fully defines the weighted two-term exponential model (\ref{eq:model3}). In each case, $\lambda_{1}$, $\lambda_{2}$ and $\omega$ are calculated by solving the boundary value problem (\ref{eq:Mk_ODE})--(\ref{eq:Mk_BC2_homog}) for $k=0,1,2$, calculating $\langle M_k(r) \rangle$ (\ref{eq:Mk_avg_homog}) for $k=0,1,2$ and then processing (\ref{eq:lambda1_weight})--(\ref{eq:omega}). Results for Cases C/D are again combined for succinctness with the formulas given for Case D only and the formulas for Case C obtained by setting $\beta_{1} = 0$.
\bigskip

\noindent\textbf{Case A:} homogeneous slab, circular or spherical geometry ($\ell_0 = 0$, $\ell_1 = L$) with radial symmetry at the origin ($[a_0,b_0]=[0,1]$) and an absorbing boundary ($[a_1,b_1]=[1,0]$)
\begin{gather}
\label{eq:model3_lambda1}
	\lambda_{1} = \frac{d(d+2)D}{L^{2}(1+\sqrt{d(1-\theta)/(\theta(d+4))})},\\
\label{eq:model3_lambda2}
	\lambda_{2} = \frac{d(d+2)D}{L^{2}(1-\sqrt{d\theta/((1-\theta)(d+4))})},\\ \omega = \frac{d+4}{d}\left[\frac{6-d}{d+6}\right]^{2}.
\end{gather}

\medskip
\noindent\textbf{Case B:} homogeneous slab, circular or spherical geometry ($\ell_0 = 0$, $\ell_1 = L$) with radial symmetry at the origin ($[a_0,b_0]=[0,1]$) and a semi-absorbing boundary ($[a_1,b_1]=[1,\beta_{1}]$)
\begin{gather*}
	\lambda_{1} = \frac{d(d+2)D}{L^{2}(1+\sqrt{d(1-\theta)/(\theta(d+4))})+\beta_1L(d+2)}, \\
	\lambda_{2} = \frac{d(d+2)D}{L^{2}(1-\sqrt{d\theta/((1-\theta)(d+4))})+\beta_1L(d+2)}, \\ \omega = \frac{d+4}{dL^{4}}\left[\frac{(6-d)L^{2} + (d+2)(d+6)\beta_1L}{d+6}\right]^{2}.
\end{gather*}

\newpage
\noindent\textbf{Case C/D:} homogeneous slab, annular or spherical shell geometry ($\ell_0 > 0$) with a reflecting inner boundary ($[a_0,b_0]=[0,1]$) and a semi-absorbing outer boundary ($[a_1,b_1]=[1,\beta_{1}]$)

\medskip
\noindent\textit{Slab $(d=1)$}
\begin{gather*}
	\lambda_1 = \frac{3 D}{(\ell _1-\ell_0)^{2}(1 + \sqrt{(1-\theta)/(5\theta)}) + 3\beta_1(\ell_{1}-\ell_{0})}, \\
	\lambda_2 = \frac{3 D}{(\ell _1-\ell_0)^{2}(1 - \sqrt{\theta/(5(1-\theta))}) + 3\beta_1(\ell_{1}-\ell_{0})}, \\
	\omega = \frac{5 (5(\ell_1 - \ell_0) + 21 \beta_1 )^2}{49 (\ell _1-\ell_0)^2}.
\end{gather*}

\medskip
\noindent\textit{Annular $(d=2)$}
\begin{gather*}
	\lambda_1=\frac{8D(\ell_{1}^{2}-\ell_{0}^{2})}{(\ell_{1}^{2}-\ell_{0}^{2})(\ell_{1}^{2}-3\ell_{0}^{2})+4\ell_{0}^{4}\log(\ell_1/\ell_0)+4\beta_{1}(\ell_{1}^{2}-\ell_{0}^{2})^{2}/\ell_{1}+\sqrt{(1-\theta)\kappa_{2,1}/(3\theta)}}, \\
	\lambda_2=\frac{8D(\ell_{1}^{2}-\ell_{0}^{2})}{(\ell_{1}^{2}-\ell_{0}^{2})(\ell_{1}^{2}-3\ell_{0}^{2})+4\ell_{0}^{4}\log(\ell_1/\ell_0)+4\beta_{1}(\ell_{1}^{2}-\ell_{0}^{2})^{2}/\ell_{1}-\sqrt{\theta\kappa_{2,1}/(3(1-\theta))}}, \\
	\omega = \frac{(288\kappa_{2,2}\log^{2}(\ell_1/\ell_0)-1152\ell_{0}^{8}\ell_{1}^{3}(\ell_{0}^{2}+\ell_{1}^{2})\log^{3}(\ell_1/\ell_0)+24\kappa_{2,3}\log(\ell_1/\ell_0) + \kappa_{2,4})}{12\ell_1^2\kappa_{2,1}^3}, \\
	\kappa_{2,1} = (\ell_{1}^{2}-\ell_{0}^{2})^{3}(\ell_{1}^{2}-7\ell_{0}^{2})-24\ell_{0}^{4}\ell_{1}^{2}\log(\ell_1/\ell_0)(2\ell_{0}^{2}\log(\ell_1/\ell_0)+\ell_{0}^{2}-\ell_{1}^{2}), \\
	\kappa_{2,2} = \ell_{0}^{6}\ell_{1}^{2}(\ell_{1}^{2}-\ell_{0}^{2})[\ell_{1}(5\text{\ensuremath{\ell_{0}^{2}}}+\ell_{1}^{2})-4\beta_{1}(\ell_{1}^{2}-\ell_{0}^{2})], \\
	\kappa_{2,3} = \ell_{0}^{4}\ell_{1}(\ell_{1}^{2}-\ell_{0}^{2})^{2}[7\ell_{0}^{4}-12\ell_{0}^{2}\ell_{1}^{2}+24\beta_{1}\ell_{1}(\ell_{1}^{2}-\ell_{0}^{2})], \\
	\kappa_{2,4} = (\ell_{1}^{2}-\ell_{0}^{2})^{3}[\ell_{1}(3\ell_{1}^{6}-25\ell_{0}^{2}\ell_{1}^{4}+83\ell_{0}^{4}\ell_{1}^{2}-145\ell_{0}^{6})-24\beta_{1}(7\ell_{0}^{2}-\ell_{1}^{2})(\ell_{1}^{2}-\ell_{0}^{2})^{2}].
\end{gather*}

\medskip
\noindent\textit{Spherical shell $(d=3)$}
\begin{gather*}
	\lambda_1=\frac{15D(\ell_{1}^{3}-\ell_{0}^{3})}{(\ell_{1}-\ell_{0})^{3}(\ell_{1}^{2}+3\ell_{0}\ell_{1}+6\ell_{0}^{2}+5\ell_{0}^{3}/\ell_{1}+\sqrt{3\kappa_{3,1}(1-\theta)/(7\theta)})+5\beta_1(\ell_{1}^{3}-\ell_{0}^{3})^{2}/\ell_{1}^{2}}, \\
	\lambda_2=\frac{15D(\ell_{1}^{3}-\ell_{0}^{3})}{(\ell_{1}-\ell_{0})^{3}(\ell_{1}^{2}+3\ell_{0}\ell_{1}+6\ell_{0}^{2}+5\ell_{0}^{3}/\ell_{1}-\sqrt{3\kappa_{3,1}\theta/(7(1-\theta))})+5\beta_1(\ell_{1}^{3}-\ell_{0}^{3})^{2}/\ell_{1}^{2}}, \\
	\omega = \frac{7(\ell _1\kappa_{3,2}(\ell_1-\ell_0)^3(\ell_1^2+4\ell_0\ell_1+10\ell_0^2)+15\kappa_{3,1}\beta_1(\ell_1^3-\ell_0^3)^2)^2}{27\kappa_{3,1}^3\ell_1^4(\ell_1-\ell_0)^6},\\
	\kappa_{3,1} = \ell_{1}^{4}+6\ell_{0}\ell_{1}^{3}+21\ell_{0}^{2}\ell_{1}^{2}+41\ell_{0}^{3}\ell_{1}+36\ell_{0}^{4}, \\
	\kappa_{3,2} = \ell_{1}^{5}+5\ell_{0}\ell_{1}^{4}+15\ell_{0}^{2}\ell_{1}^{3}+50\ell_{0}^{3}\ell_{1}^{2}+100\ell_{0}^{4}\ell_{1}+54\ell_{0}^{5}.
\end{gather*}

\medskip
\noindent\textbf{Case E:} homogeneous slab, annular or spherical shell geometry ($\ell_0 > 0$) with absorbing inner ($[a_0,b_0]=[1,0]$) and outer ($[a_1,b_1]=[1,0]$) boundaries

\medskip
\noindent\textit{Slab $(d=1)$}
\begin{gather*}
	\lambda_1 = \frac{12D}{(\ell_1 - \ell_0)^2(1+\sqrt{(1-\theta)/(5\theta)})}, \\
	\lambda_2 = \frac{12D}{(\ell_1 - \ell_0)^2(1-\sqrt{\theta/(5(1-\theta))})}, \\
	\theta = \frac{1}{2} + \frac{1}{2}\sqrt{125/321}.
\end{gather*}

\medskip
\noindent\textit{Annular $(d=2)$}
\begin{gather*}
	\lambda_{1}=\frac{8D\log(\ell_1/\ell_0)}{(\ell_{0}^{2}+\ell_{1}^{2})\log(\ell_1/\ell_0)-(\ell_{1}^{2}-\ell_{0}^{2})+\sqrt{(1-\theta)\xi_{2,1}/(3\theta)}}, \\
	\lambda_{2}=\frac{8D\log(\ell_1/\ell_0)}{(\ell_{0}^{2}+\ell_{1}^{2})\log(\ell_1/\ell_0)-(\ell_{1}^{2}-\ell_{0}^{2})-\sqrt{\theta\xi_{2,1}/(3(1-\theta))}}, \\
	\omega =\frac{\log^{2}(\ell_1/\ell_0)(18(\ell_{0}^{2}+\ell_{1}^{2})(\ell_{1}^{2}-\ell_{0}^{2})^{2}+6(\ell_{0}^{2}+\ell_{1}^{2})(\ell_{0}^{4}+\ell_{1}^{4})\log^{2}(\ell_1/\ell_0)-\xi_{2,2}\log(\ell_1/\ell_0))^{2}}{48\xi_{2,1}^{3}}, \\
	\xi_{2,1} = 3(\ell_{1}^{2}-\ell_{0}^{2})^{2}-3(\ell_{1}^{4}-\ell_{0}^{4})\log(\ell_1/\ell_0)+(\ell_{1}^{2}-\ell_{0}^{2})^{2}\log^{2}(\ell_1/\ell_0), \\
	\xi_{2,2} =(\ell_{1}^{2}-\ell_{0}^{2})(19\ell_{0}^{4}+46\ell_{0}^{2}\ell_{1}^{2}+19\ell_{1}^{4}).
\end{gather*}

\medskip
\noindent\textit{Spherical shell $(d=3)$}
\begin{gather*}
	\lambda_{1}=\frac{60D(\ell_{1}^{3}-\ell_{0}^{3})}{(\ell_{1}-\ell_{0})^{3}(4\ell_{0}^{2}+7\ell_{0}\ell_{1}+4\ell_{1}^{2}+\sqrt{3(1-\theta)\xi_{3,1}/(7\theta)})}, \\
	\lambda_{2}=\frac{60D(\ell_{1}^{3}-\ell_{0}^{3})}{(\ell_{1}-\ell_{0})^{3}(4\ell_{0}^{2}+7\ell_{0}\ell_{1}+4\ell_{1}^{2}-\sqrt{3\theta\xi_{3,1}/(7(1-\theta))})}, \\
	\omega = \frac{7(64\ell_{0}^{6}+471\ell_{0}^{5}\ell_{1}+780\ell_{0}^{4}\ell_{1}^{2}+745\ell_{0}^{3}\ell_{1}^{3}+780\ell_{0}^{2}\ell_{1}^{4}+471\ell_{0}\ell_{1}^{5}+64\ell_{1}^{6})^{2}}{27\xi_{3,1}^{3}}, \\
	\xi_{3,1} = 16\ell_{1}^{4}+26\ell_{0}\ell_{1}^{3}+21\ell_{0}^{2}\ell_{1}^{2}+26\ell_{0}^{3}\ell_{1}+16\ell_{0}^{4}.
\end{gather*}

\bigskip
\noindent The above results yield easy-to-evaluate surrogate models that provide analytical insight into the role of dimension, diffusivity, geometry and boundary conditions on the proportion of particles remaining over time, $\mathcal{P}(t)$. For Case A, the expressions for $\lambda_{1}$ (\ref{eq:model3_lambda1}) and $\lambda_{2}$ (\ref{eq:model3_lambda2}) in the weighted two-term exponential model take a similar form to the expression for $\lambda_{1}$ and $\lambda_{2}$ in the two-term exponential model (\ref{eq:model2_case1}), with the exception of correction terms in the denominator depending on the weighting $\theta$. Using these expressions for $\lambda_{1}$ and $\lambda_{2}$, we see that the weighted two-term exponential model exhibits initial decay rates of $\theta\lambda_{1}+(1-\theta)\lambda_{2} = 10D/L^2, 24D/L^2, 42D/L^2$ for $d = 1,2,3$, each of which exceed the initial decay rate of the two-term exponential model.

\section{Results}\label{sec:results}
We now investigate the accuracy of the three surrogate models (\ref{eq:c1}), (\ref{eq:model2}) and (\ref{eq:model3}). Here, we consider the seven test cases outlined previously in \hyperlink{tab1}{Table 1} but with specific choices for the parameters as detailed in \hyperlink{tab2}{Table 2}. For the homogeneous geometries (Cases A--E), we choose $P=\delta=\tau=1$ giving $D = P\delta^2/(2d\tau) = 1/(2d)$ while for the heterogeneous geometries (Cases F--G) we choose $P_1=0.3$ and $P_2=\delta=\tau=1$ giving $D_1=P_1\delta^2/(2d\tau)=0.3/(2d)$ and $D_2=P_2\delta^2/(2d\tau)=1/(2d)$. The surrogate models for Cases A--E are given in sections \ref{sec:one_term}--\ref{sec:two_term_weight} while the surrogate models for Cases F--G are given in sections \ref{sec:one_term}--\ref{sec:two_term}. Surrogate model parameter values for Cases A--G in either one ($d=1$), two ($d=2$) or three ($d=3$) dimensions can be found in \hyperlink{appxA}{Appendix A}. All simulations are performed over a specified time interval $0<t<T=2\log(10)/\lambda$, where $\lambda$ is the decay rate in the one-term exponential model (section \ref{sec:one_term}). This choice of $T$ corresponds to the value of time satisfying $S_1(t) = 10^{-2}$ and captures the main region of decay of $\mathcal{P}(t)$ to easily detect differences between the surrogate models. 

\begin{table}[H]\hypertarget{tab2}{}
\centering
\setlength{\tabcolsep}{0.45em}
\begin{tabular}{c l c c c p{2.7cm} p{4.5cm} c c c c}
	\hline
	Case & Geometry & $\ell_0$ & $\ell_1$ & $\ell_2$ & Inner Boundary & Outer Boundary & $a_0$ & $b_0$ & $a_1$ & $b_1$ \\
	\hline
	A & homogeneous & 0 & 100 & -- & -- & absorbing & 0 & 1 & 1 & 0 \\
	B & homogeneous & 0 & 100 & -- & -- & semi-absorbing ($P_{\rm{O}} = 0.5$) & 0 & 1 & 1 & 2 \\
	C & homogeneous & 50 & 100 & -- & reflecting & absorbing & 0 & 1 & 1 & 0 \\
	D & homogeneous & 50 & 100 & -- & reflecting & semi-absorbing ($P_{\rm{O}} = 0.5$) & 0 & 1 & 1 & 2 \\
	E & homogeneous & 50 & 100 & -- & absorbing & absorbing & 1 & 0 & 1 & 0 \\
	F & heterogeneous & 0 & 50 & 100 & -- & absorbing & 0 & 1 & 1 & 0 \\
	G & heterogeneous & 0 & 50 & 100 & -- & semi-absorbing ($P_{\rm{O}} = 0.5$) & 0 & 1 & 1 & 2 \\
	\hline
\end{tabular}
	\caption{Geometry and boundary parameters for Cases A--G.}
\end{table}

Each surrogate model is benchmarked against the stochastic and continuum model. To account for the variability of $\mathcal{P}_s(t)$ (\ref{eq:avg_stoch}) from the stochastic model, we perform $N_s = 100$ simulations using $N_p=50$ and $N_p=500$ particles. For each value of $N_p$, we store the minimum and maximum values of $\mathcal{P}_s(t)$ at each time step across all $N_s = 100$ simulations with the resulting area enclosed encompassing all realizations of $\mathcal{P}_s(t)$. For the heterogeneous geometries (Cases F--G), we choose $n = 36$ and $n_{1}=n_{2}=12$ \cite{carr_2020} when processing the movement probabilities at the interface (see section \ref{sec:SRW_HETERO}). To calculate $\mathcal{P}_c(t)$ from the continuum model, we first compute a numerical solution to the homogeneous continuum model (\ref{eq:PDE_HOMOG})--(\ref{eq:BC2_HOMOG}) (Cases A--E) or the heterogeneous continuum model (\ref{eq:PDE_HETERO})--(\ref{eq:IF2}) (Cases F--G) by discretising in space using a finite volume method and discretising in time using the Crank-Nicolson method. We use $N_t=10^{5}$ fixed time steps and $N_{r} = 501$ (Cases A--E) or $N_r = 1001$ (Cases F--G) uniformly-spaced nodes. For both the homogeneous continuum model (\ref{eq:PDE_HOMOG})--(\ref{eq:BC2_HOMOG}) and the heterogeneous continuum model (\ref{eq:PDE_HETERO})--(\ref{eq:IF2}), this yields approximations $c(r_i,t_j)$ where $r_i = \ell_0 + (i-1)(\ell_m-\ell_0)/(N_r-1)$ ($m = 1$ for Cases A--E and $m=2$ for Cases F--G) and $t_j = jT/N_t$ for $i=1,\hdots,N_r$ and $j=1,\hdots,N_t$. Using these discrete approximations, $c(r_i,t_j)$, and a trapezoidal rule approximation to the integrals (\ref{eq:avg_homog}) or (\ref{eq:avg_hetero}) then allows $P_{c}(t_{j})$ to be computed for $j=1,\hdots,N_t$. In addition to visual comparisons, to quantify the accuracy of the surrogate models, we also use the mean absolute error between each surrogate model and $\mathcal{P}_c(t)$,
\begin{gather}\label{eq:meanErrors}
	\varepsilon_k = \frac{1}{N_t}\sum_{j=1}^{N_t} |C_k(t_j) - \mathcal{P}_c(t_j)|,
\end{gather}
where $k=1,\hdots,3$. Full details of the above implementations are available in our MATLAB code which can be accessed on GitHub: \href{https://github.com/lukefilippini/Filippini\_2023}{https://github.com/lukefilippini/Filippini\_2023}.

\begin{figure}[p]\hypertarget{fig2}{}
\centering
\includegraphics[width=1.0\textwidth]{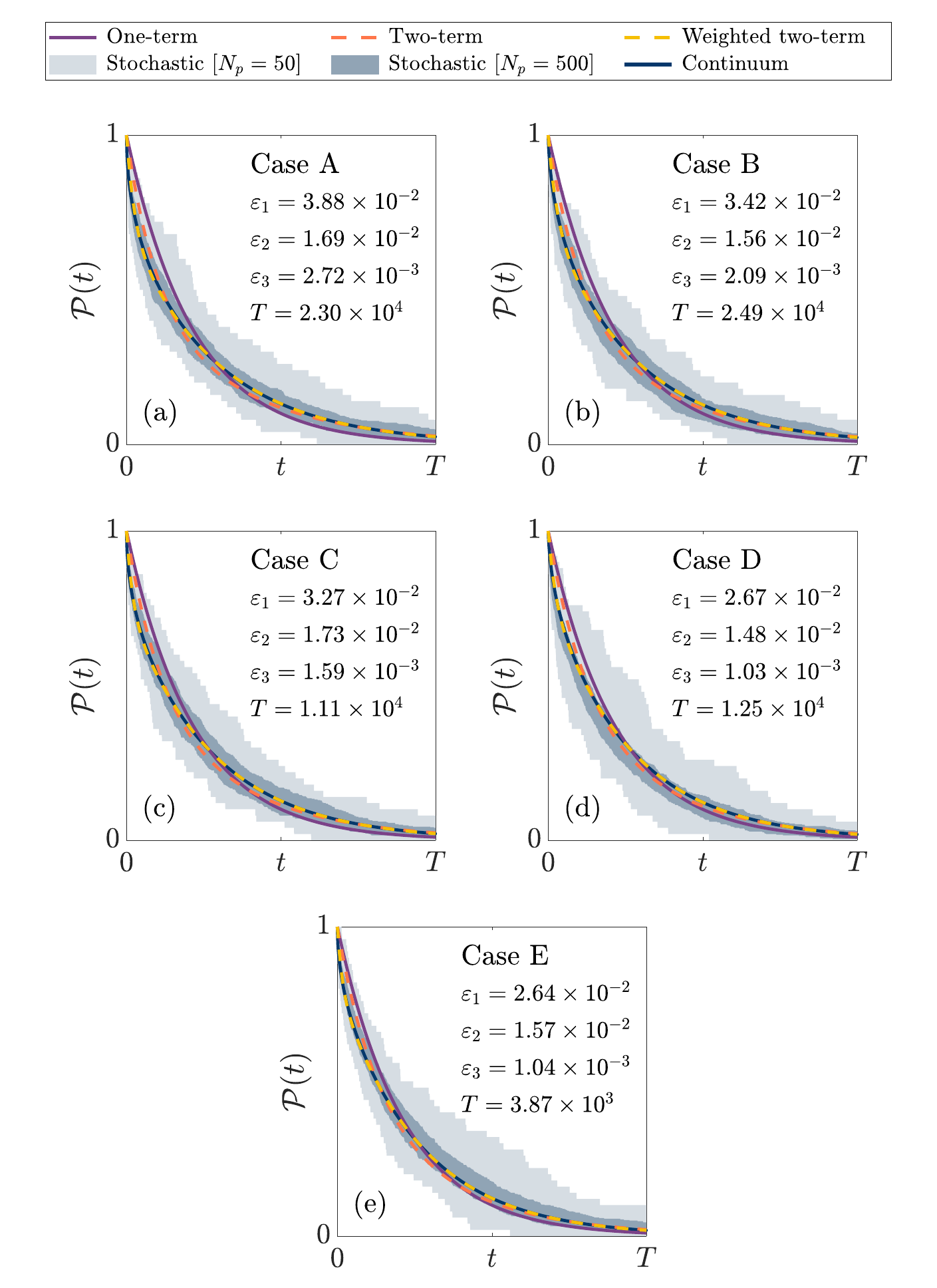}
\caption{One-term, two term, and weighted two-term exponential models for $\mathcal{P}(t)$ compared with stochastic and continuum models for the homogeneous test cases (Cases A--E) with $d = 2$. For the stochastic model, the bounds of the shaded regions represent the maximum and minimum proportion of particles remaining at each point in time across the $N_s=100$ simulations. The mean absolute errors $\varepsilon_1$, $\varepsilon_2$ and $\varepsilon_3$ and final time $T$ are rounded to three significant digits.}
\end{figure}
\begin{figure}[t]\hypertarget{fig3}{}
\centering
\includegraphics[width=1.0\textwidth]{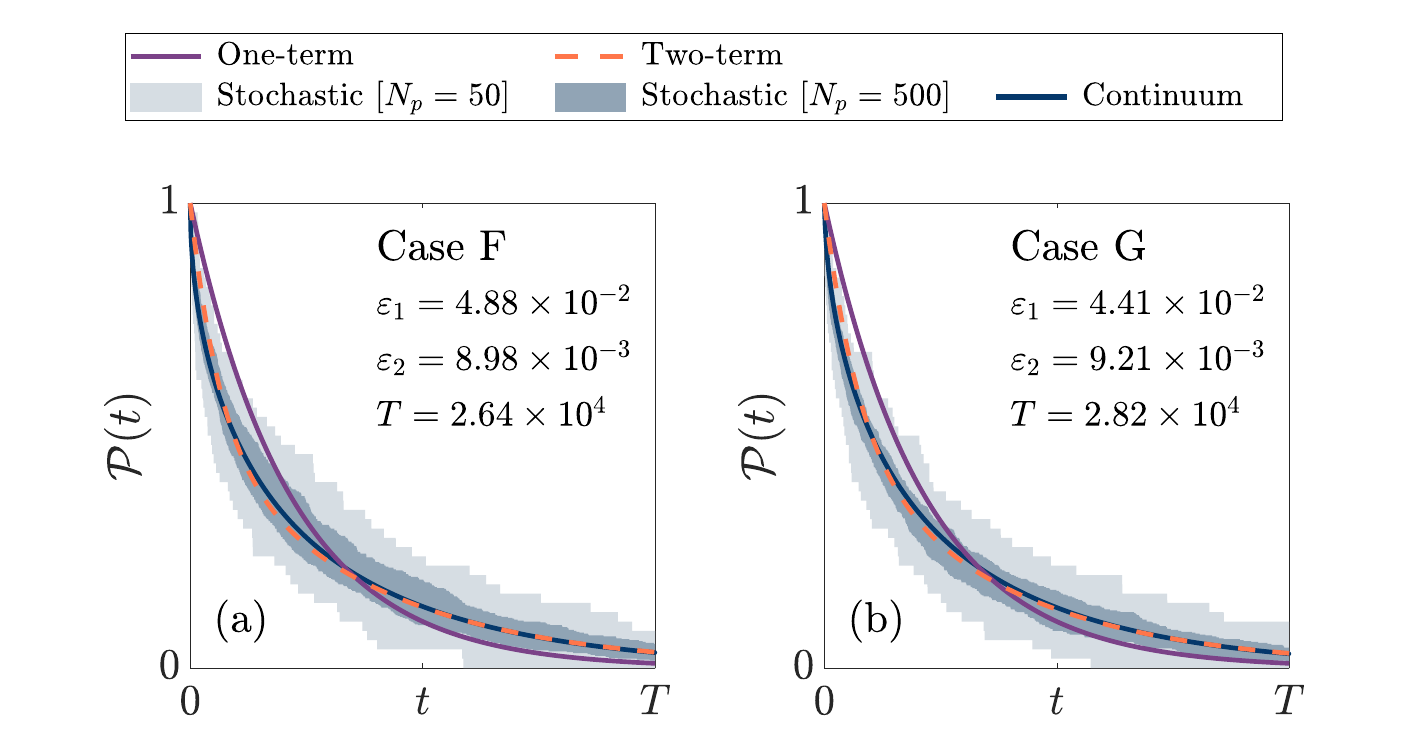}
\caption{One-term and two-term exponential models for $\mathcal{P}(t)$ compared with stochastic and continuum models for the heterogeneous test cases (Cases F--G) with $d = 2$. For the stochastic model, the bounds of the shaded regions represent the maximum and minimum proportion of particles remaining at each point in time across the $N_s=100$ simulations. The mean absolute errors $\varepsilon_1$ and $\varepsilon_2$ and final time $T$ are rounded to three significant digits.}
\end{figure}

\hyperlink{fig2}{Figure 2} and \hyperlink{fig3}{Figure 3} compare the surrogate models  to the benchmark values of $\mathcal{P}_{s}(t)$ and $\mathcal{P}_{c}(t)$ obtained from the stochastic and continuum models. \hyperlink{fig2}{Figure 2} assesses the performance of the one-term (\ref{eq:c1}), two-term (\ref{eq:model2}) and weighted two-term exponential models (\ref{eq:model3}) for the homogeneous test cases (Cases A--E) while \hyperlink{fig3}{Figure 3} assesses the performance of the one-term (\ref{eq:c1}) and two-term (\ref{eq:model2}) exponential models for the heterogeneous test cases (Cases F--G). All subfigures feature the final time $T$ and the corresponding mean absolute errors (\ref{eq:meanErrors}) for each surrogate model. Results are shown for $d=2$ only with similar results observed for $d=1,3$. From the results in \hyperlink{fig2}{Figure 2} and \hyperlink{fig3}{Figure 3}, we can conclude that:
\begin{itemize}
	\item All surrogate models reliably capture the release profile over the seven test cases but with varying levels of accuracy.
	\item The one-term exponential model (\ref{eq:c1}) has the lowest accuracy of the three models across all seven test cases, however, it is the most simplistic and may potentially be sufficient in some cases.
	\item The weighted two-term exponential model (\ref{eq:model3}) provides the highest accuracy, capturing the early and late decay of $\mathcal{P}(t)$ more accurately than the one-term (\ref{eq:c1}) and two-term (\ref{eq:model2}) models, however, this comes with the trade-off of increased model complexity.
	\item All three surrogate models yield higher accuracy for test cases with semi-absorbing boundary conditions (Cases B, D and G) when compared to test cases with purely absorbing boundary conditions (Cases A, C, E and F).
\end{itemize}

\noindent Finally, we compare the surrogate models developed in this paper to the Weibull model of Carr \cite{carr_2022}, which was developed for homogeneous geometries only (i.e. Cases A--E). \hyperlink{weibull}{Table 3} displays the mean absolute errors for the Weibull model, denoted as $\varepsilon_w$, for Cases A--E. In comparison to the mean absolute errors presented in \hyperlink{fig2}{Figure 2}, we find that the Weibull model is more accurate than the one-term (\ref{eq:c1}) and two-term (\ref{eq:model2}) exponential models but less accurate than the weighted two-term exponential model (\ref{eq:model3}).

\begin{table}[H]\hypertarget{weibull}{}
	\centering
	\begin{tabular}{c c c c c c}
		\hline
		Case &  A & B & C & D & E \\
		\hline
		$\varepsilon_w $ & $1.05 \times 10^{-2}$ & $9.21 \times 10^{-3}$ & $1.03 \times 10^{-2}$ & $8.42 \times 10^{-3}$ & $9.22 \times 10^{-3}$ \\
		\hline
	\end{tabular}
	\caption{Mean absolute errors for the Weibull model of Carr \cite{carr_2022} (Cases A--E).}
\end{table}

\section{Conclusion}\label{sec:conclusion}
We have considered the problem of particle diffusion in $d$-dimensional radially-symmetric geometries with reflecting, absorbing and/or semi-absorbing boundaries. By matching moments with the continuum analogue of the stochastic diffusion model, we have presented several new one-term and two-term exponential models for $\mathcal{P}(t)$, the proportion of particles remaining within the geometry over time. New surrogate models have been developed for three main problems: (i) homogeneous slab, circular and spherical geometries with an absorbing or semi-absorbing outer boundary (ii) homogeneous slab, annular and spherical shell geometries with absorbing, reflecting and/or semi-absorbing boundaries and (ii) heterogeneous slab, circular and spherical geometries with an absorbing or semi-absorbing outer boundary. Each surrogate model provides a simple approximation of $\mathcal{P}(t)$ that is easy to evaluate, avoids the limitations and complexity of exact expressions obtained from the continuum model, reliably captures the particle release profile over time and explicitly depends on the physical parameters of the diffusive transport system: dimension, diffusivity, geometry and boundary conditions. Of the three surrogate models developed, our findings demonstrated that the weighted two-term exponential model (\ref{eq:model3}) captures both stochastic and continuum calculations of $\mathcal{P}(t)$ with the highest degree of accuracy. It also offers improved simplicity and accuracy when compared to the Weibull model previously presented by Carr \cite{carr_2022}. 

The results reported in this paper indicate, as may have been expected, that the most accurate surrogate model is the one with the greatest number of parameters (weighted two-term exponential model) and the least accurate surrogate model is the one with the fewest number of parameters (one-term exponential model). To account for this trade-off between model accuracy and model complexity, a standard model selection criterion that rewards accuracy but penalises the number of parameters \cite{simpson_2022} could be used to select a single preferred surrogate model. Other avenues for future work could include accounting for a non-uniform initial distribution of particles or drift and/or decay in the diffusive transport process. Surrogate models using different functional forms, such as a weighted two-term Weibull model, could also be explored and may further improve accuracy but at the cost of increased model complexity. 

\section*{Acknowledgements}
This research was supported by an Australian Government Research Training Program (RTP) scholarship, which provided LPF with a stipend during his Master of Philosophy candidature.


\section*{Declaration of competing interest}
The authors declare that they have no known competing financial interests or personal relationships that could have appeared to influence the work in this paper.

\section*{Data availability}
All supporting code is available on GitHub: \href{https://github.com/lukefilippini/Filippini_2023}{https://github.com/lukefilippini/Filippini\_2023}.

\appendix

\setcounter{table}{0}
\renewcommand{\thetable}{A\arabic{table}}

\section*{Appendix A. Surrogate model parameter values}\hypertarget{appxA}{}
Here, we present the one-term (\ref{eq:c1}), two-term (\ref{eq:model2}) and weighted two-term (\ref{eq:model3}) exponential models corresponding to Cases A--G (see \hyperlink{tab2}{Table~2}) by providing numerical values for the parameters appearing in each surrogate model. All values are rounded to three significant figures and presented in tables separated by dimension. 

\label{sec:tables}
\begin{table}[H]
	\centering
	\begin{tabular}{c c c c c c c}
		\hline
		& \multicolumn{1}{c}{One-term} & \multicolumn{2}{c}{Two-term} & \multicolumn{3}{c}{Two-term (weighted)} \\
		\hline
		Case & $\lambda$ & $\lambda_1$ & $\lambda_2$ & $\lambda_1$ & $\lambda_2$ & $\theta$ \\
		\hline
		A & $1.50 \times 10^{-4}$ & $1.04 \times 10^{-4}$ & $2.71 \times 10^{-4}$ & $1.23 \times 10^{-4}$ & $2.13 \times 10^{-4}$ & $8.12 \times 10^{-1}$ \\
		B & $1.42 \times 10^{-4}$ & $9.95 \times 10^{-5}$ & $2.45 \times 10^{-4}$ & $1.19 \times 10^{-4}$ & $1.84 \times 10^{-3}$ & $8.27 \times 10^{-1}$ \\
		C & $6.00 \times 10^{-4}$ & $4.15 \times 10^{-4}$ & $1.09 \times 10^{-3}$ & $4.94 \times 10^{-4}$ & $8.51 \times 10^{-3}$ & $8.12 \times 10^{-1}$ \\
		D & $5.36 \times 10^{-4}$ & $3.83 \times 10^{-4}$ & $8.92 \times 10^{-4}$ & $4.56 \times 10^{-4}$ & $6.57 \times 10^{-3}$ & $8.41 \times 10^{-1}$ \\
		E & $2.40 \times 10^{-3}$ & $1.66 \times 10^{-3}$ & $4.34 \times 10^{-3}$ & $1.96 \times 10^{-3}$ & $3.40 \times 10^{-2}$ & $8.12 \times 10^{-1}$ \\ 
		F  & $1.16 \times 10^{-4}$ & $7.34 \times 10^{-5}$ & $2.79 \times 10^{-4}$ & -- & -- & -- \\
		G & $1.11 \times 10^{-4}$ & $7.13 \times 10^{-5}$ & $2.51 \times 10^{-3}$ & -- & -- & -- \\ 
		\hline 
	\end{tabular}
	\caption{Surrogate model parameters for Cases A--G appearing in the one-term (\ref{eq:c1}), two-term (\ref{eq:model2}) and weighted two-term (\ref{eq:model3}) exponential models ($d=1$).}
\end{table}

\begin{table}[H]
\centering
\begin{tabular}{c c c c c c c}
	\hline
	& \multicolumn{1}{c}{One-term} & \multicolumn{2}{c}{Two-term} & \multicolumn{3}{c}{Two-term (weighted)} \\
	\hline
	Case & $\lambda$ & $\lambda_1$ & $\lambda_2$ & $\lambda_1$ & $\lambda_2$ & $\theta$ \\
	\hline
	A & $2.00 \times 10^{-4} $ & $1.27 \times 10^{-4}$ & $4.73 \times 10^{-4}$ & $1.45 \times 10^{-4}$ & $1.65 \times 10^{-3}$ & $6.99 \times 10^{-1}$ \\
	B & $1.85 \times 10^{-4}$ & $1.21 \times 10^{-4}$ & $3.98 \times 10^{-4}$ & $1.39 \times 10^{-4}$ & $1.39 \times 10^{-3}$ & $ 7.24 \times 10^{-1} $ \\
	C & $4.16 \times 10^{-4}$ & $2.75 \times 10^{-4} $ & $8.55 \times 10^{-4}$ & $3.22 \times 10^{-4}$ & $4.58 \times 10^{-3}$ & $7.58 \times 10^{-1}$ \\
	D & $3.70 \times 10^{-4}$ & $2.54 \times 10^{-4}$ & $6.80 \times 10^{-4}$ & $3.00 \times 10^{-4}$ & $3.50 \times 10^{-3}$ & $7.93 \times 10^{-1}$ \\
	E & $1.19 \times 10^{-3}$ & $8.20 \times 10^{-4}$ & $2.17 \times 10^{-3}$ & $9.76 \times 10^{-4}$ & $1.61 \times 10^{-2}$ & $8.08 \times 10^{-1}$ \\ 
	F & $1.75 \times 10^{-4}$ & $1.02 \times 10^{-4}$ & $6.09 \times 10^{-4}$ & -- & -- & -- \\
	G & $1.63 \times 10^{-4}$ & $9.79 \times 10^{-5}$ & $4.90 \times 10^{-4}$ & -- & -- & -- \\
	\hline
\end{tabular}
	\caption{Surrogate model parameters for Cases A--G  appearing in the one-term (\ref{eq:c1}), two-term (\ref{eq:model2}) and weighted two-term (\ref{eq:model3}) exponential models ($d=2$).}
\end{table}

\begin{table}[H]
\centering
\begin{tabular}{c c c c c c c}
	\hline
	& \multicolumn{1}{c}{One-term} & \multicolumn{2}{c}{Two-term} & \multicolumn{3}{c}{Two-term (weighted)} \\
	\hline
	Case & $\lambda$ & $\lambda_1$ & $\lambda_2$ & $\lambda_1$ & $\lambda_2$ & $\theta$ \\
	\hline
	A & $2.50 \times 10^{-4}$ & $1.51 \times 10^{-4}$ & $7.24 \times 10^{-4}$ & $1.66 \times 10^{-4}$ & $1.58 \times 10^{-3}$ & $6.23 \times 10^{-1}$ \\
	B & $2.27 \times 10^{-4}$ & $1.42 \times 10^{-4}$ & $5.61 \times 10^{-4}$ & $1.59 \times 10^{-4}$ & $1.29 \times 10^{-3}$ & $6.57 \times 10^{-1}$ \\
	C & $3.78 \times 10^{-4}$ & $2.40 \times 10^{-4}$ & $8.95 \times 10^{-4}$ & $2.75 \times 10^{-4}$ & $3.34 \times 10^{-3}$ & $7.03 \times 10^{-1}$ \\
	D & $3.34 \times 10^{-4}$ & $2.21 \times 10^{-4}$ & $6.81 \times 10^{-4}$ & $2.57 \times 10^{-4}$ & $2.52 \times 10^{-3}$ & $7.44 \times 10^{-1}$ \\
	E & $8.24 \times 10^{-4}$ & $5.57 \times 10^{-4}$ & $1.58 \times 10^{-3}$ & $6.59 \times 10^{-4}$ & $9.58 \times 10^{-3}$ & $7.85 \times 10^{-1}$ \\ 
	F  & $2.33 \times 10^{-4}$ & $1.32 \times 10^{-4}$ & $9.86 \times 10^{-4}$ & -- & -- & -- \\
	G & $2.13 \times 10^{-4}$ & $1.26 \times 10^{-4}$ & $6.98 \times 10^{-4}$ & -- & -- & -- \\ 
	\hline 
\end{tabular}
	\caption{Surrogate model parameters for Cases A--G  appearing in the one-term (\ref{eq:c1}), two-term (\ref{eq:model2}) and weighted two-term (\ref{eq:model3}) exponential models ($d=3$).}
\end{table}

\bibliography{references}

\begin{thebibliography}{10}

\bibitem{codling_2008}
E.~A. Codling, M.~J. Plank, and S.~Benhamou.
\newblock Random walk models in biology.
\newblock {\em Journal of the Royal Society Interface}, 5:813--834, 2008.

\bibitem{lotstedt_2015}
P.~Lötstedt and L.~Meinecke.
\newblock Simulation of stochastic diffusion via first exit times.
\newblock {\em Journal of Computational Physics}, 300:862--886, 2015.

\bibitem{okubo_2001}
A.~Okubo and S.~A. Levin.
\newblock {\em Diffusion and ecological problems: Modern perspectives}.
\newblock Springer, New York, 2nd edition, 2001.

\bibitem{arifin_2006}
D.~Y. Arifin, L.~Y. Lee, and C.~H. Wang.
\newblock Mathematical modeling and simulation of drug release from
  microspheres: Implications to drug delivery systems.
\newblock {\em Advanced Drug Delivery Reviews}, 58(12--13):1274--1325, 2006.

\bibitem{dash_2010}
S.~Dash, P.~N. Murthy, L.~Nath, and P.~Chowdhury.
\newblock Kinetic modeling on drug release from controlled drug delivery
  systems.
\newblock {\em Acta Poloniae Pharmaceutica}, 67(3):217--223, 2010.

\bibitem{siepmann_book_2012}
J.~Siepmann, R.~A. Siegel, and M.~J. Rathbone.
\newblock {\em Fundamentals and applications of controlled release drug
  delivery}.
\newblock Springer, New York, 2012.

\bibitem{vaccario_2015}
G.~Vaccario, C.~Antoine, and J.~Talbot.
\newblock First-passage times in $d$-dimensional heterogeneous media.
\newblock {\em Physical Review Letters}, 115(24), 2015.

\bibitem{van_kampen_2007}
N.~G. v.~Kampen.
\newblock {\em Stochastic processes in physics and chemistry}.
\newblock Elsevier, Amsterdam, 3rd edition, 2007.

\bibitem{jain_2022}
A.~Jain, S.~{McGinty}, G.~Pontrelli, and L.~Zhou.
\newblock Theoretical modeling of endovascular drug delivery into a multilayer
  arterial wall from a drug-coated balloon.
\newblock {\em International Journal of Heat and Mass Transfer}, 187:122572,
  2022.

\bibitem{siepmann_2012}
J.~Siepmann and F.~Siepmann.
\newblock Modeling of diffusion controlled drug delivery.
\newblock {\em Journal of Controlled Release}, 161(2):351--362, 2012.

\bibitem{carr_2018}
E.~J. Carr and G.~Pontrelli.
\newblock Modelling mass diffusion for a multi-layer sphere immersed in a
  semi-infinite medium: Application to drug delivery.
\newblock {\em Mathematical Biosciences}, 303:1--9, 2018.

\bibitem{kaoui_2018}
B.~Kaoui, M.~Lauricella, and G.~Pontrelli.
\newblock Mechanistic modelling of drug release from multi-layer capsules.
\newblock {\em Computers in Biology and Medicine}, 93:149--157, 2018.

\bibitem{aghbashlo_2009}
M.~Aghbashlo, M.~H. Kianmehr, S.~Khani, and M.~Ghasemi.
\newblock Mathematical modelling of thin-layer drying of carrot.
\newblock {\em International Agrophysics}, 23(4):313--317, 2009.

\bibitem{corzo_2008}
O.~Corzo, N.~Bracho, A.~Pereira, and A.~Vásquez.
\newblock Weibull distribution for modeling air drying of coroba slices.
\newblock {\em LWT -- Food Science and Technology}, 41(10):2023--2028, 2008.

\bibitem{midilli_2002}
A.~Midilli, H.~Kucuk, and Z.~Yapar.
\newblock A new model for single-layer drying.
\newblock {\em Drying Technology}, 20(7):1503--1513, 2002.

\bibitem{onwude_2016}
D.~I. Onwude, N.~Hashim, R.~B. Janius, N.~M. Nawi, and K.~Abdan.
\newblock Modeling the thin-layer drying of fruits and vegetables: A review.
\newblock {\em Comprehensive Reviews in Food Science and Food Safety},
  15(3):599--618, 2016.

\bibitem{redner_2001}
S.~Redner.
\newblock {\em A guide to first-passage processes}.
\newblock Cambridge University Press, New York, 2001.

\bibitem{simpson_2015}
M.~J. Simpson and R.~E. Baker.
\newblock Exact calculations of survival probability for diffusion on growing
  lines, disks, and spheres: The role of dimension.
\newblock {\em Journal of Chemical Physics}, 143(9), 2015.

\bibitem{carr_2022}
E.~J. Carr.
\newblock Exponential and {Weibull} models for spherical and spherical-shell
  diffusion-controlled release systems with semi-absorbing boundaries.
\newblock {\em Physica A: Statistical Mechanics and its Applications},
  605:127985, 2022.

\bibitem{ignacio_2017}
M.~Ignacio, M.~V. Chubynsky, and G.~W. Slater.
\newblock Interpreting the {Weibull} fitting parameters for
  diffusion-controlled release data.
\newblock {\em Physica A: Statistical Mechanics and its Applications},
  486:486--496, 2017.

\bibitem{andrews_2016}
C.~J. Andrews, L.~Cuttle, and M.~J. Simpson.
\newblock Quantifying the role of burn temperature, burn duration and skin
  thickness in an in vivo animal skin model of heat conduction.
\newblock {\em International Journal of Heat and Mass Transfer}, 101:542--549,
  2016.

\bibitem{simpson_2009}
M.~J. Simpson.
\newblock Depth-averaging errors in reactive transport modeling.
\newblock {\em Water Resources Research}, 45(2):W02505, 2009.

\bibitem{ignacio_2022}
M.~Ignacio, M.~Bagheri, M.~V. Chubynsky, H.~W. de~Haan, and G.~W. Slater.
\newblock Diffusivity interfaces in lattice {Monte Carlo} simulations: Modeling
  inhomogeneous delivery and release systems.
\newblock {\em Physical Review E}, 105(6):064135, 2022.

\bibitem{gomes_filho_2020}
M.~S. Gomes-Filho, M.~A.~A. Barbosa, and F.~A. Oliveira.
\newblock A statistical mechanical model for drug release: Relations between
  release parameters and porosity.
\newblock {\em Physica A: Statistical Mechanics and its Applications},
  540:123165, 2020.

\bibitem{carr_2020}
E.~J. Carr, J.~M. Ryan, and M.~J. Simpson.
\newblock Diffusion in heterogeneous discs and spheres: New closed-form
  expressions for exit times and homogenization formulas.
\newblock {\em Journal of Chemical Physics}, 153(7):074115, 2020.

\bibitem{erban_2007}
R.~Erban and S.~J. Chapman.
\newblock Reactive boundary conditions for stochastic simulations of
  reaction-diffusion processes.
\newblock {\em Physical Biology}, 4(1):16--28, 2007.

\bibitem{crank_1975}
J.~Crank.
\newblock {\em The mathematics of diffusion}.
\newblock Oxford University Press, New York, 2nd edition, 1975.

\bibitem{simon_2016}
L.~Simon and J.~Ospina.
\newblock {\em Closed-form solutions for drug transport through
  controlled-release devices in two and three dimensions}.
\newblock John Wiley \& Sons, Inc, New Jersey, 2016.

\bibitem{carr_2016}
E.~J. Carr and I.~W. Turner.
\newblock A semi-analytical solution for multilayer diffusion in a composite
  medium consisting of a large number of layers.
\newblock {\em Applied Mathematical Modelling}, 40:7034--7050, 2016.

\bibitem{hickson_2009}
R.~I. Hickson, S.~I. Barry, and G.~N. Mercer.
\newblock Critical times in multilayer diffusion. part 1: Exact solutions.
\newblock {\em International Journal of Heat and Mass Transfer}, 52:5776--5783,
  2009.

\bibitem{ignacio_2021}
M.~Ignacio and G.~W. Slater.
\newblock Using fitting functions to estimate the diffusion coefficient of drug
  molecules in diffusion-controlled release systems.
\newblock {\em Physica A: Statistical Mechanics and its Applications},
  567:125681, 2021.

\bibitem{carr_2019}
E.~J. Carr and M.~J. Simpson.
\newblock New homogenization approaches for stochastic transport through
  heterogeneous media.
\newblock {\em The Journal of Chemical Physics}, 150(4):044104, 2019.

\bibitem{simpson_2022}
M.~J. Simpson, A.~P. Browning, D.~J. Warne, O.~J. Maclaren, and R.~E. Baker.
\newblock Parameter identifiability and model selection for sigmoid population
  growth models.
\newblock {\em Journal of Theoretical Biology}, 535:110998, 2022.

\end{thebibliography}
\bibliographystyle{unsrt}

\end{document}